\documentclass[preprint]{aastex6}

\usepackage{amsmath}
\usepackage{bm}
\usepackage{tabularx}

\usepackage{setspace}
\usepackage{amssymb}	
\usepackage{multirow}
\usepackage{morefloats}

\usepackage{natbib}

\begin{document}

\title{Neptune's resonances in the Scattered Disk}

\author{Lei Lan\altaffilmark{1}, Renu Malhotra\altaffilmark{2}}
\affil{The University of Arizona}

\altaffiltext{1}{Also Tsinghua University, lanl14@mails.tsinghua.edu.cn}
\altaffiltext{2}{Corresponding author; renu@lpl.arizona.edu}

\shortauthors{Lan \& Malhotra}

\begin{abstract}
The Scattered Disk Objects (SDOs) are thought to be a small fraction of the ancient population of leftover planetesimals in the outer solar system that were gravitationally scattered by the giant planets and have managed to survive primarily by capture and sticking in Neptune's exterior mean motion resonances (MMRs). In order to advance understanding of the role of MMRs in the dynamics of the SDOs, we investigate the phase space structure of a large number of Neptune's MMRs in the semi-major axis range 33--140~au by use of Poincar\'e sections of the circular planar restricted three body model for the full range of particle eccentricity pertinent to SDOs. We find that, for eccentricities corresponding to perihelion distances near Neptune's orbit, distant MMRs have stable regions with widths that are surprisingly large and of similar size to those of the closer-in MMRs. We identify a phase-shifted second resonance zone that exists in the phase space at planet-crossing eccentricities but not at lower eccentricities; this second resonance zone plays an important role in the dynamics of SDOs in lengthening their dynamical lifetimes. Our non-perturbative measurements of the sizes of the stable resonance zones  confirm previous results and provide an additional explanation for the prominence of the $N$:1 sequence of MMRs over the $N$:2, $N$:3 sequences and other MMRs in the population statistics of SDOs;  our results also provide a tool to more easily identify resonant objects. 
\end{abstract}


\section{Introduction} \label{sec:intro}

The population of near- and beyond-Neptune minor planets discovered over the past two-and-a-half decades provides important diagnostics about the origin and evolution of our solar system \citep{Luu:2002, Morbidelli:2004}.  A number of these objects are recognized to be part of an extensive ``scattered'' and ``scattering'' population that has incurred gravitational scattering encounters with Neptune over its past history; this population inhabits a region of orbital parameter space defined by somewhat fuzzy boundaries of semimajor axis $a$ in the range $30~{\rm au}\lesssim a \lesssim 2000~{\rm au}$ and perihelion distance $q$ in the range $5~{\rm au}\lesssim q\lesssim$~37~au~\citep{Gladman:2008}. Theoretical models suggest that the present-day Scattered Disk Objects (SDOs) are just a small, $\sim1\%$, fraction of the ancient population of leftover planetesimals formed in the outer solar system that were gravitationally scattered by the giant planets~\citep{Levison:1997,Duncan:1997,Lykawka:2007a,Gomes:2008}. The SDOs are postulated to be a major source of the transient populations of Centaurs and the Jupiter-family comets and possibly also contribute to the Halley-type comets and the Oort Cloud \citep{Duncan:1997, Fernandez:2004, Emelyanenko:2005, Levison:2006, Volk:2008}.

\begin{figure*}
 \centering
 \includegraphics[width=190mm]{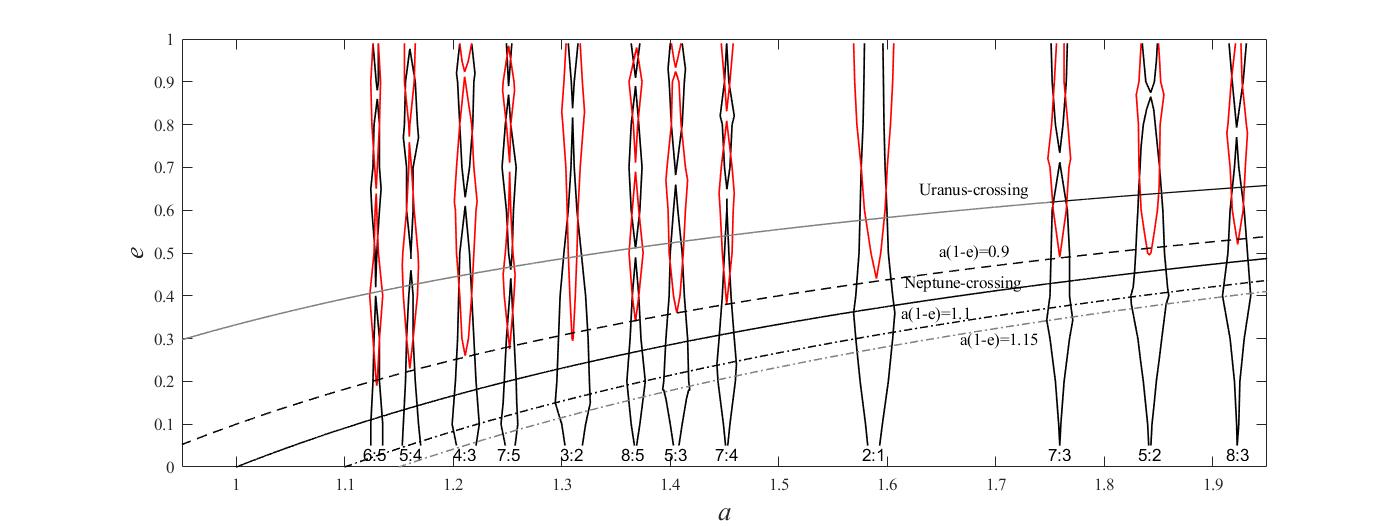}
 \caption{ 
Stable libration zones, in the $(a,e)$ plane, for a selection of Neptune's exterior MMRs. The semimajor axis $a$ is in units of Neptune's semimajor axis, $a_{\rm{Neptune}}=30.1$~au. The area bounded by the black lines is the width in semi-major axis for the first resonance zone, and the area bounded by the red lines is that of the second resonance zone (see Section~\ref{s:Poincare} for details). For reference, we also draw several curves of constant perihelion distance, $a(1-e)=$~constant: the uppermost curve (in continuous gray line) indicates orbits of perihelion distance equal to Uranus' aphelion distance; the dashed black curve indicates perihelion distance equal to 0.9 $a_{\rm{Neptune}}$ $(\sim27.1$~au); the continuous black curve indicates perihelion distance equal to Neptune's semimajor axis; the dot-dash curves in black and grey indicate perihelion distances of 1.1 $a_{\rm{Neptune}}$ $(\sim33.1$~au) and 1.15 $a_{\rm{Neptune}}$ $(\sim34.6$~au), respectively. Scattered disk objects diffuse chaotically mostly in the perihelion distance range 30--36~AU while often sticking near the resonance boundaries.}
 \label{f:f1b}
\end{figure*}

The median dynamical lifetime of a minor planet subsequent to a strong gravitational scattering encounter with Neptune is about ten million years~\citep{Levison:1997,Tiscareno:2003,diSisto:2007}.  Typically, such encounters result in further strong scattering by the other giant planets, leading to either ejection of the small body from the solar system or to collisions with the planets or the Sun. The survival and persistence of a remnant of the ancient planetesimal disk in the form of the present-day Scattered Disk population over the age of the solar system is currently understood to be owed to a phenomenon called ``resonance sticking" which ensures that when these objects approach Neptune's orbit near their perihelion their true longitude is usually well-separated from Neptune's~\citep{Gomes:2008}. This is not dissimilar to the well-understood mechanism for the long term stability of Pluto's Neptune-crossing orbit~\citep{Milani:1989,Malhotra:1997}. However, unlike the stable libration of Pluto's perihelion well away from Neptune's true longitude, the SDOs evolve near the chaotic boundaries of MMRs such that they are not strictly prohibited from close encounters with Neptune. Numerical models find that the SDOs have only occasional gravitational scattering encounters with Neptune but otherwise spend a large fraction of their dynamical lifetime in the vicinity of mean motion resonances with that planet~\citep{Duncan:1997,Lykawka:2007b,Volk:2018,Yu:2018}.  In this ``resonance sticking" behavior, the SDOs evolve chaotically in semimajor axis and eccentricity and visit many different MMRs for varying random lengths of time while maintaining their perihelion distance mostly in the narrow range of 30--35 au (see Figure~\ref{f:f1b}).  The simulations indicate that, in order of decreasing importance, the $N$:1, $N$:2 and $N$:3 resonances play the most important roles in the resonance sticking phenomenon, in terms of both capture probabilities and the residence time in the MMRs.  This pattern for the SDOs' resonance sticking is consistent with a prior theoretical study by \cite{Pan:2004} who showed that, in the simple model of the circular planar restricted three body problem, the $N$:1 exterior resonances have the largest resonance widths when the Jacobi constant is close to 3; the $N$:2, $N$:3, etc.~resonances have decreasing size of resonance widths when the particle eccentricity is in the scattering regime. However, it is in contrast with the classical perturbation theoretic analyses of mean motion resonances at lower eccentricities which anticipates that resonance width decreases exponentially with the order $|q|$ of a $p:p+q$ resonance~\citep{Murray:1999SSD}. In the high eccentricity (scattering) regime, \cite{Pan:2004} suggested a redefinition of the ``order" of resonance as ``$p$"-th order for a $p:p+q$ resonance.

The ``resonance sticking" mechanism is complex and we still have only limited understanding of it. A number of previous studies have examined the dynamics of Neptune's MMRs with semi-analytical and numerical models in various approximations~\citep[e.g.][]{Beauge:1994,Morbidelli:1995,Malhotra:1996,Morbidelli:1997,Nesvorny:2000,Nesvorny:2001,Robutel:2001,Chiang:2002,Kotoulas:2004,Kotoulas:2005,Voyatzis:2005,Voyatzis:2005b,Murray-Clay:2005, Gallardo:2006b,Tiscareno:2009,Saillenfest:2016,Saillenfest:2017b,Yu:2018,Gallardo:2018,Gallardo:2019}. 
The present paper provides another view on Neptune's exterior resonances pertinent to understanding their role in the resonance sticking dynamics of the SDOs.
We begin by recognizing that, in the high eccentricity regime of the SDOs, the chaotic separatrices at the boundaries of the stable resonance zones are undoubtedly important in the ``resonance sticking" dynamics but their role in defining the widths of resonances is not accurately computed in analyses that isolate a single resonance and/or average over short period perturbations.
 The nature of the phase space is revealed more clearly with non-perturbative numerical analyses of the behavior of planet-crossing orbits.
We therefore investigate the phase space structure of a wide range of Neptune's MMRs by examination of Poincar\'e surfaces of section in the circular planar restricted three body model of the Sun, Neptune and a test particle, with the latter representing an SDO.  The advantage of this simple model is that its two dimensional Poincar\'e sections serve to accurately compute and visualize the chaotic boundaries of resonances and provide a non-perturbative measure of the widths of the stable libration zones near resonances for the full range of particle eccentricity. A disadvantage is that non-co-planarity, the effects of a non-circular Neptune orbit, and the perturbations of other planets are not included; still, the map of stable libration zones of moderate-to-high-eccentricity MMRs in the simplified model yields new insights, as we will show.  
In the context of the outer solar system, this non-perturbative method has been previously employed by
\cite{Pan:2004} for their study of exterior resonant orbits of Jacobi constant close to 3 (pericenter distance close to the planet's orbit); \cite{Wang:2017} employed it for a general study of the 2:1 and 3:2 interior resonances in the high eccentricity regime.
\cite{Malhotra:1996} and \cite{Malhotra:2018} made use of this method for a small number of Neptune's exterior MMRs. 
Here we use this method for a large number of Neptune's exterior MMRs pertinent to the dynamics of SDOs, in the semi-major axis range $34-150$~au (particle-to-Neptune period ratio from 6:5 to 10:1) and test particle eccentricity in the range $0.05$ to $\sim1$. The lower end of this eccentricity range is somewhat arbitrary but sufficient to encompass the moderate-to-high eccentricity regime pertinent to the scattered and scattering disk. 
We carry out comparisons of different resonances to identify patterns that can help towards a better understanding of the resonance sticking phenomenon.  Additionally, because our investigation yields fairly accurate boundaries of the libration zones of MMRs in the semi-major axis--eccentricity $(a,e)$ plane, a by-product of our investigation is a new tool to identify resonant minor planets beyond Neptune; this tool is complementary to the current method of arduous examination of the behavior of critical resonant angles in $\sim10$~megayear long numerical integrations~\citep[e.g.,][]{Gladman:2008,Volk:2016,Forgacs-Dajka:2018}.

\section{Poincar\'e sections for Neptune's exterior resonances} \label{s:Poincare}

To study the phase space structure and dynamics near Neptune's mean motion resonances, we generated Poincar\'e sections of the circular planar restricted three body model of the Sun, Neptune and a massless test particle, with the latter representing a Kuiper belt object.
Because Neptune's orbit is nearly circular (its eccentricity does not exceed $\sim0.025$ over long timescales~\citep{Murray:1999SSD}), and because the masses of the Kuiper belt objects are very small ($\lesssim10^{-4}$ of Neptune's), this simplified model is sufficient to identify the basic phase structure and dynamics near Neptune's exterior mean motion resonances.
This model neglects the influence of Neptune's eccentricity and the solar system planets interior to Neptune, as well as non-coplanar motion, but affords the advantage of visualization of the phase space structure in two-dimensional sections.  These neglected effects are not critical for the present purpose. In a study of Neptune's 5:2 MMR, \cite{Malhotra:2018} showed that perturbations of the other giant planets (Jupiter, Saturn and Uranus) and non-co-planarity exert only a mild influence on the boundaries of the resonance zone in the $(a,e)$ plane. 

Natural units are adopted for this model: we set the unit of mass to be $m_1+m_2$ (where $m_1$ and $m_2$ denote the masses of Sun and Neptune, respectively), the unit of length to be the constant orbital separation of the two primaries (i.e., Neptune's semi-major axis, $a_{\rm Neptune}$), and the unit of time to be their orbital period divided by $2\pi$. With these units, Newton's constant of gravitation and the orbital angular velocity of the two primaries about their barycenter are unity. We employed the equations of motion for the test particle in the rotating reference frame of constant unit angular velocity with origin at the barycenter of $m_1$ and $m_2$; in this frame, the two primaries are fixed at locations, $(-\mu,0)$ and $(1-\mu,0)$, respectively, where $\mu=m_1/(m_1+m_2)=5.146\times10^{-5}$ is the fractional mass of Neptune.  For the numerical integrations of the equations of motion we used the adaptive step size seventh-order Runge-Kutta method~\citep{Fehlberg:1968}, with relative and absolute error tolerances of $10^{-12}$.

We adopted a systematic strategy for the exploration of near-resonant particle initial conditions as in \cite{Malhotra:2018}. 
Briefly, all the test particle trajectories represented in a Poincar\`e section have the same value of the Jacobi integral and the same initial values of semi-major axis and eccentricity, but different values of the initial longitude of perihelion.  
The particle's initial osculating semi-major axis is always set equal to the nominal resonant value, $a_{\rm{res}}=(N/k)^{2/3}$, where $N$ and $k$ are mutually prime integers, and $N/k>1$ is the period ratio of test particle and Neptune. The initial location of each particle in its orbit is always at its perihelion but different particles have different initial pericenter longitudes spanning the full range $(0^\circ,360^\circ$). We integrate the particle motion for several thousand orbital periods, recording its state vector ($x,y,\dot{x},\dot{y}$) at every perihelion passage. We then transform the state vector into osculating orbital elements and generate the Poincar\'e section as a plot of ($\psi,a$), where $a$ is the osculating barycentric semi-major axis, and $\psi$ is the true longitude separation between Neptune and the test particle when the particle is at perihelion.  
The phase angle $\psi$ is related to the usual critical resonant angle $\phi$ for an $N$:$k$ resonance as follows: 
\begin{equation}
\phi \equiv k\,\lambda_{\rm Neptune}-N\lambda+(N-k)\,\varpi 
 \quad = k\,\psi . \label{e:phi}
\end{equation}  

Figures~\ref{f:2t1} and \ref{f:3t2} show examples of Poincar\'e sections for Neptune's 2:1 and 3:2 exterior MMRs. 
In these plots, we see stable resonant orbits that make closed smooth curves and we also see chaotic orbits that wander over a wider area and do not stay confined to smooth curves. 
The region containing a sequence of closed smooth curves is a stable resonance island.  The center of a stable island indicates the exact resonant orbit (corresponding to a periodic orbit), whereas the smooth curves surrounding it are orbits librating about the exact resonance (corresponding to quasi-periodic orbits). 
The centers of the stable resonant islands are located at a few discrete values of $\psi$, each with a semi-major axis value close to $a_{\rm{res}}$.  
Each Poincar\'e section is labeled with the value of the initial eccentricity common to all the trajectories represented in that section; this value is very close to the eccentricity of the periodic orbit at the center of the stable islands in that section. 
For each initial eccentricity value, we measure the width of the stable island by its range in semi-major axis, $\Delta a = a_{\rm{max}}-a_{\rm{min}}$, where $a_{\rm{max}}$ and $a_{\rm{min}}$ are the upper and lower semi-major axis boundaries of the stable island. For most MMRs, the chaotic orbits are not visible at low eccentricities, but become visible at moderate and high eccentricity, as scattered points filling an area bounding the stable islands.  As illustrated in Figures~\ref{f:2t1} and \ref{f:3t2}, the phase space structure changes with eccentricity, with at least one particularly significant transition when new stable islands appear when eccentricity exceeds the Neptune-crossing value.  We show in Sections 4 and 5 that these phase space transitions are related to the shape of the resonant orbit in the rotating frame and its relationship to Neptune's orbit (see also \cite{Wang:2017} and \cite{Malhotra:2018}).

After generating many Poincar\'e sections and measuring the stable resonance boundaries, $a_{\rm{max}}$ and $a_{\rm{min}}$, for the full range of eccentricities for many MMRs, we gather the results and plot the resonance boundaries in the $(a,e)$ parameter plane. Figures \ref{f:f1b}, \ref{f:nt1} and \ref{f:nt2} show the boundaries of resonances with the particle-to-Neptune period ratios from 6:5 to 10:1. Each MMR has two distinct libration zones, indicated as the ``first resonance zone" (black curves) and the ``second resonance zone" (red curves). These two zones have stable islands that are centered at different values of the angular separation, $\psi$, of the particle's perihelion longitude relative to Neptune. The stable island centers of the first resonance zone avoid $\psi=0$, whereas the second resonance zone has a stable island centered at $\psi=0$.  Physically this means that the periodic orbits at the center of each of these zones correspond to traces in the rotating frame that have the same shape but different geometrical orientations relative to the Sun-Neptune line. In terms of the critical resonant angle, $\phi$ (Eq.~\ref{e:phi}), the first resonance zone has librations of $\phi$ centered at $\pi$ whereas the second resonance zone has librations of $\phi$ centered at 0. In the case of the $N$:1 resonances, the first resonance zone has two additional libration centers displaced on either side of $\phi=\pi$, as can be seen in the top panels of Figure~\ref{f:2t1} for the 2:1 MMR; the location of these additional libration centers varies with eccentricity (see Section~\ref{subsec:PatternsNto1}). 

For all the MMRs investigated here, the first resonance zone exists over almost the entire range of eccentricity; for some resonances, it disappears for a small range of high eccentricities, then reappears again at an even higher eccentricity. The second resonance zone emerges at an eccentricity slightly exceeding the critical value, $e_c$, for Neptune-crossing, 
\begin{equation}
e_c=1-a_{\rm{Neptune}}/a_{\rm res}.
\label{e:ec}
\end{equation} 
For the $N$:1 resonances, the stable islands of the second resonance zone expand monotonically with increasing eccentricity, but this is not the case for the $N$:2, the $N$:3 and higher resonances; for latter, the second resonance zone has several transitions with increasing eccentricity, first expanding then shrinking, even disappearing (e.g., the 3:2, 5:3 and 8:5 resonances).

\section{\it{N}\rm:1 resonances} \label{sec:nto1}

We computed the Poincar\'e sections for the $N$:1 sequence of Neptune's exterior MMRs, from 2:1 to 10:1, for the range of particle eccentricities 0.05--1. The 2:1 MMR is the closest-in of this sequence, and its phase space structure is exemplary of all the $N$:1 sequence of MMRs.  Below we describe this MMR in some detail.

\subsection{Neptune's 2:1 Exterior MMR}

\begin{figure*}
 \centering
 \includegraphics[width=170mm]{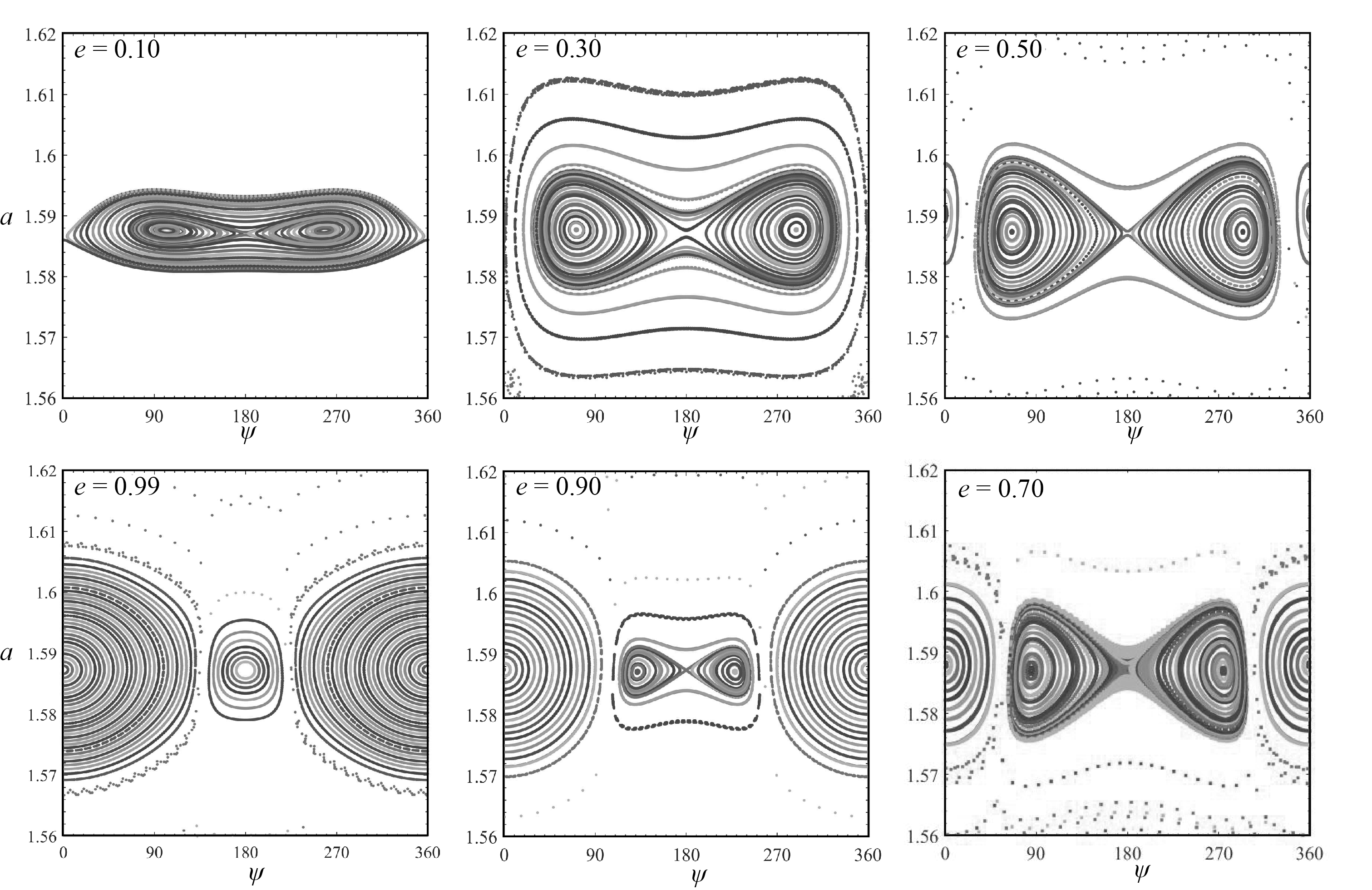}
 \caption{
The surfaces of section in ($\psi,a$) near Neptune's exterior 2:1 resonance, at different particle eccentricities. Clockwise from top left: $e=0.10$, $e=0.30$, $e=0.50$, $e=0.70$, $e=0.90$, $e=0.99$. In the first two panels, only the first resonance zone is visible; it consists of the asymmetric and symmetric librators. In the remaining panels there is another resonant island visible, centered at $\psi=0$; this is the second resonance zone.
}
 \label{f:2t1}
\end{figure*}

\begin{figure*}
 \centering
 \includegraphics[width=140mm]{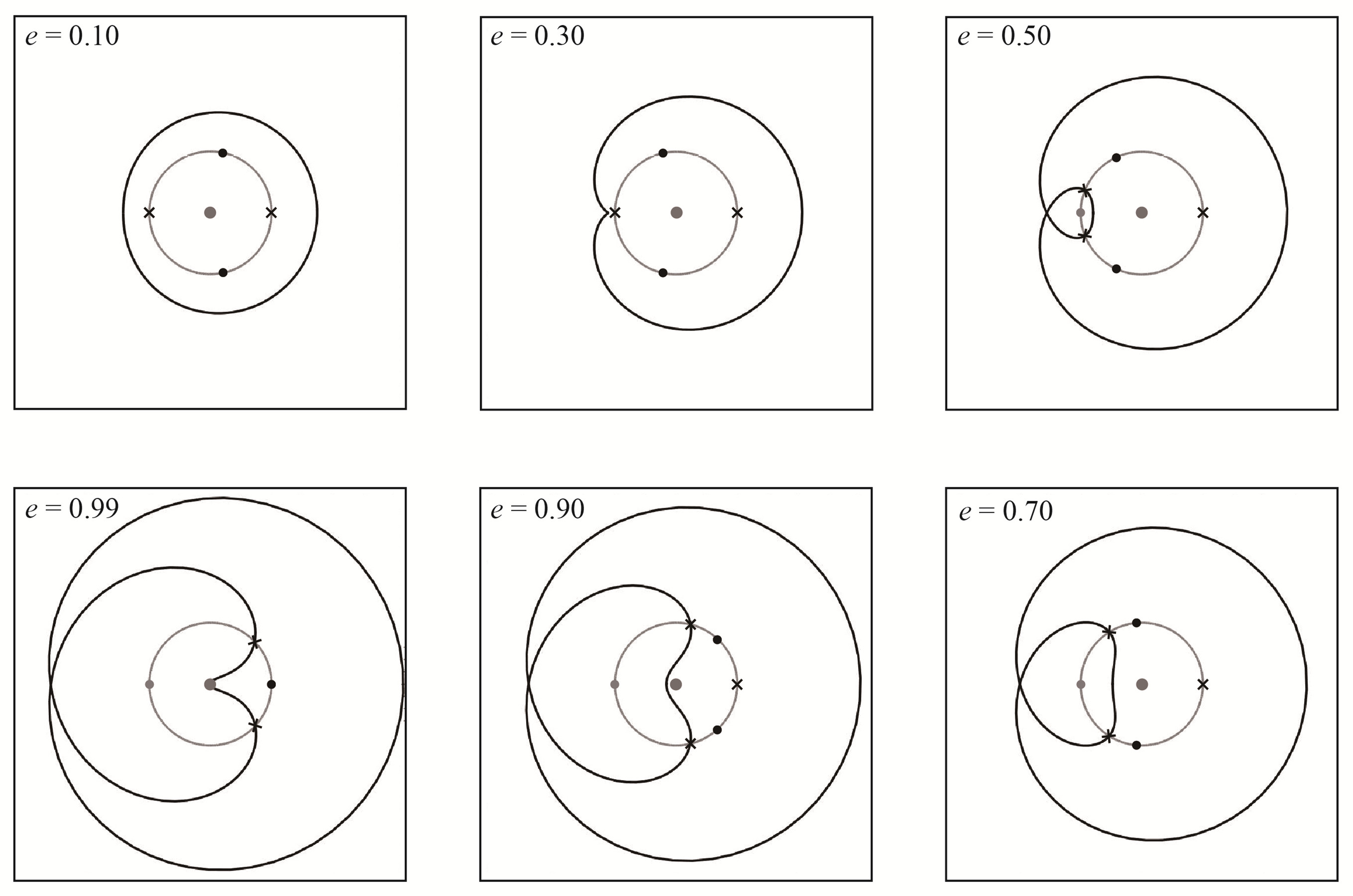}
 \caption{
 The geometry of the exterior 2:1 resonant orbit in the rotating frame. The black curve is the trace of the particle's eccentric orbit; note that there is only one pericenter passage represented in this trace.  On the gray circle of radius $1-\mu$, we plot with a dot the (fixed) planet location corresponding to stable configurations (viz.~the centers of libration visible in the Poincar\'e sections in Figure~\ref{f:2t1}); the black dots are for the first resonance zone, the gray dots are for the second resonance zone.  Also on the gray circle are black crosses indicating either the unstable fixed points visible in the Poincar\'e sections or the collision points.  The larger gray dot near the center of the figure indicates the location of the Sun, which is always a unit distance from Neptune; the small mass ratio, $\mu=5.146\times10^{-5}$ means that the slightly different locations of the Sun for different locations of Neptune remain unresolved in these plots.}
 \label{f:21e}
\end{figure*}

\begin{figure}
 \centering
 \includegraphics[width=90mm]{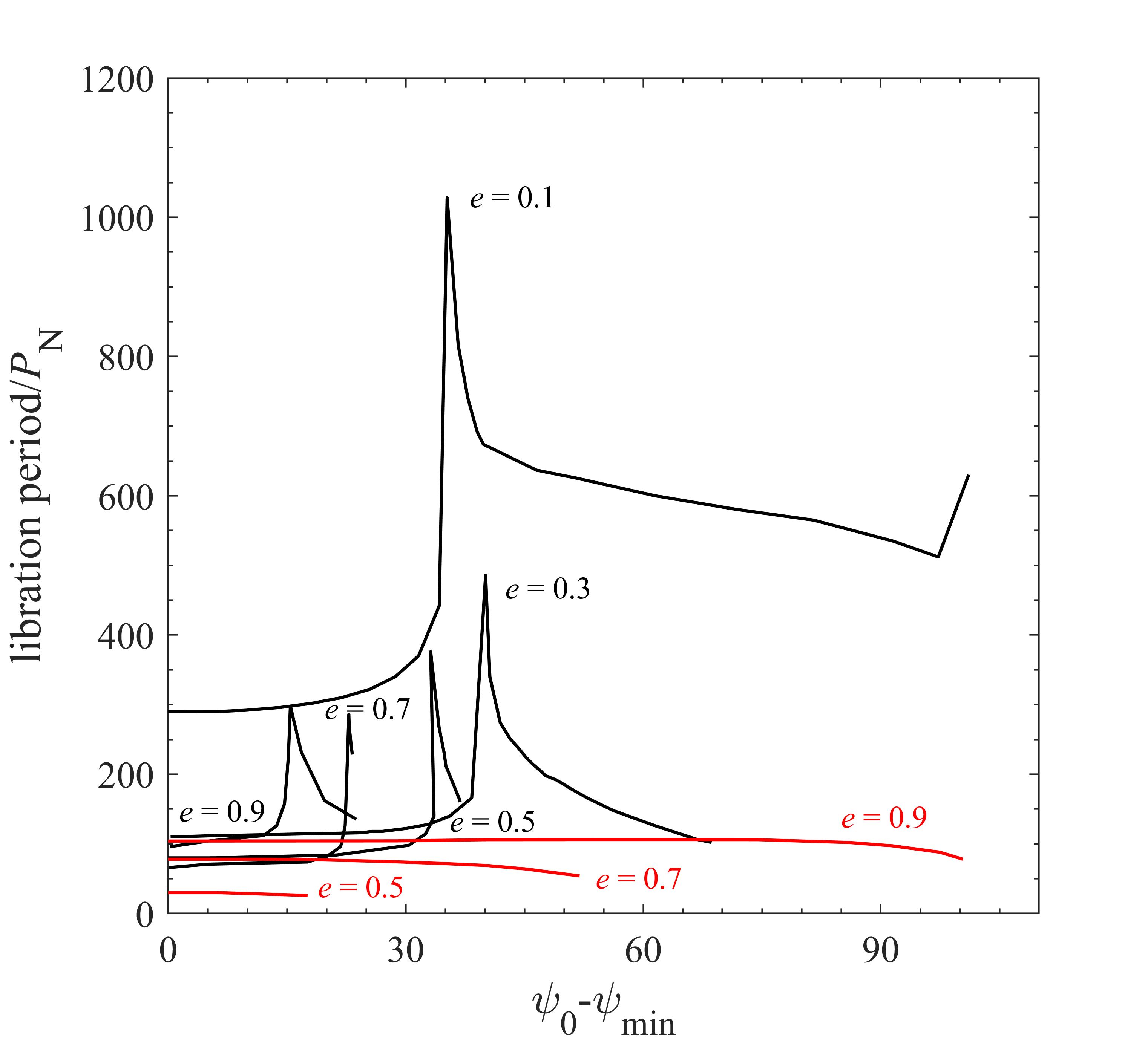}
 \caption{
The libration period (in units of Neptune's orbital period) in Neptune's exterior 2:1 resonance as a function of libration amplitude at different particle eccentricities; black curves indicate the first resonance zone and red curves indicate the second resonance zone.  }
 \label{f:21pe}
\end{figure}

Figure \ref{f:2t1} plots several representative surfaces of section in the neighborhood of Neptune's 2:1 exterior MMR at semi-major axes around $a_{\rm{res}} = (2/1)^{2/3}=1.5874$. 
We observe that at the lower value of eccentricity, near $\sim0.1$, only smooth closed curves are present, and there is no visible chaotic region.  The libration zone has semi-major axis centered at $a=1.588$, and it contains three families of stable orbits, each of which has different centers for $\psi$: a pair of families confined to small libration amplitudes has $\psi$ librating about a center near $\psi_0\approx-102^\circ$ and $\psi_0\approx+102^\circ$ ($\psi_0$ is eccentricity-dependent and varies from near $\pm180^\circ$ to near $\pm60^\circ$ -- see Fig.~\ref{f:cve}), and the third family has larger amplitude librations of $\psi$ centered at $180^\circ$.
 The family of large amplitude librations centered at $\psi=180^\circ$ is called the ``symmetric librators", and the pair of families of small amplitude librations with centers displaced to either side of $\psi=180^\circ$ are called the ``asymmetric librators"~\citep{Beauge:1994,Malhotra:1996,Nesvorny:2001,Chiang:2002,Pan:2004}.  (The boundary, or separatrix, between the asymmetric and symmetric librations is a curve shaped like the outline of a butterfly.) This nomenclature is also descriptive of the center of librations of the critical resonant angle, $\phi=\lambda_{\rm Neptune}-2\lambda+\varpi$. Physically, it is descriptive of the orientation of the particle's perihelion relative to Neptune (see Fig.~\ref{f:21e}, top left): for the symmetric libration center, the particle's perihelion longitude is $180^\circ$ away from Neptune and the trace of its orbit in the rotating frame is symmetric relative to the line joining the fixed locations of the Sun and Neptune; for the asymmetric libration centers, the angular separation of the particle's perihelion from Neptune is approximately $102^\circ$ (for particle eccentricity $\sim0.1$) and the trace of its orbit in the rotating frame is not symmetric relative to the line joining the fixed locations of the Sun and Neptune.  As evident in the surface of section, Figure~\ref{f:2t1} (top panels), the two families of asymmetric librators are confined to small libration amplitudes; the symmetric librators surround the two zones of the asymmetric librators and are accordingly confined to large libration amplitudes. 
At very low eccentricities only the symmetric librations centered at $\psi=180^\circ$ are present; for Neptune's exterior 2:1 MMR, the two families of asymmetric librators first appear at $e\simeq0.04$ \citep{Malhotra:1996}.
 We report the width in semi-major axis, $\Delta a$, of the first resonance zone as measured by the boundaries of the symmetric libration zone; it includes within it the asymmetric libration zone.

At larger eccentricities, the test particle's perihelion is closer to Neptune's orbit.  The Poincar\'e sections show that the stable island is larger in semi-major axis width, $\Delta a$. 
The center $\psi_0$ of the asymmetric librations drifts further away from $180^\circ$, and their maximum libration amplitude also increases, squeezing the libration amplitudes of the symmetric librators to a smaller range.  As seen in Figure~\ref{f:2t1}, for $e=0.30$, the orbits near the outer boundary of the libration zone are visibly chaotic. When the eccentricity exceeds the critical Neptune-crossing value, $e_c=0.370$, the particle's perihelion is interior to Neptune's orbit. At eccentricity $e_c$, we find that the width, $\Delta a$, of the libration zone reaches a maximum.
At a somewhat higher eccentricity, $e_2=0.440$, a new stable island, which is centered at $\psi_0=0^\circ$, becomes visible in the Poincar\'e section; this can be seen in Figure \ref{f:2t1} for $e=0.50$.  We call this the second resonance zone.  For increasing eccentricity above $e_2$, the stable island of the second resonance zone expands at the expense of the size of the islands of the first resonance zone (Figure \ref{f:2t1}, bottom panels).  The emergence of a new stable island signals the existence of a pair of new periodic orbits in the 2:1 exterior resonance.  The trace of the periodic orbits in the rotating frame has the same shape for the new and old periodic orbits, but their orientation relative to the fixed locations of the Sun and Neptune in the rotating fame are actually different. This is illustrated in Figure~\ref{f:21e} which shows the paths of the particle's 2:1 resonant orbit in the rotating frame, at different particle eccentricities.  We observe that, for $e>e_c$, the resonant orbit intersects itself and creates a ``perihelion lobe". The circle of unit radius about the Sun (which is the circular orbit of Neptune) cuts the perihelion lobe of the particle's trajectory.  For the second resonance zone, the fixed location of Neptune in the rotating frame is inside the perihelion lobe, whereas for the resonant orbits of the first resonance zone, Neptune's location is outside the perihelion lobe.  We observe that the length of arc of the unit circle which is interior to the perihelion lobe correlates with the range of librations of $\psi$ in the second resonance zone: the longer the length of this arc, the wider the stable islands of the second resonance zone. At an eccentricity of $\sim0.681$, the widths, $\Delta a$ in semi-major axis, of the new and old stable islands are nearly equal. For eccentricities above $0.681$, the semi-major axis width of the new stable island exceeds that of the old one, and the sizes of the libration islands of the first resonance zone shrink rapidly. The two asymmetric islands of the first resonance zone merge into a single, symmetric island when the eccentricity exceeds $\sim0.95$ (Figure \ref{f:2t1}, bottom left panel).

As noted above, the centers of the two asymmetric islands drift away from $180^\circ$ as the eccentricity increases from near 0.05 to $e_2=0.440$, and then drift back toward $180^\circ$ for eccentricity increasing from $e_2$ to $e_3=0.95$.  This behavior is shown in Figure~\ref{f:cve}.  \cite{Nesvorny:2001} have previously reported the locations of the asymmetric centers of the 2:1 resonance for eccentricity values up to 0.7; our results are similar for that case but extend to eccentricity values up to unity. We additionally report the locations of the asymmetric centers of the 3:1 and 4:1 resonances.
 (A numerical method for computing the location of the stable and unstable equilibrium points of all resonances from the resonant disturbing function is given by \cite{Gallardo:2006b, Gallardo:2019}.) For increasing $N$ of the $N$:1 resonances, it is interesting to note that the asymmetric islands first become visible at a higher threshold eccentricity.

A dynamical quantity of interest is the libration period of resonant orbits.  A librating orbit is not at exact resonance so it is not a periodic orbit; it can be thought of as tracing almost the same shape as the exact resonant orbit, but slowly drifting and librating about the exact periodic orbit.  We measured the libration periods by measuring the time for successive points in the Poincar\'e section to complete one circuit around the center of a resonant island.  (This measure is therefore only accurate to $\pm1$ orbital period of the test particle; applications desiring a more accurate measure could use the numerical filtering and frequency analysis method of \cite{Gallardo:1997}.) For Neptune's 2:1 exterior resonance, we note that each smooth closed curve in a Poincar\'e section is uniquely identified by the minimum value, $\psi_{\rm{min}}$, on a librating orbit. These libration periods are plotted as a function of libration amplitude, $\psi_0-\psi_{\rm{min}}$, in Figure \ref{f:21pe}.
In the first resonance zone, the libration period is not a monotonic function of eccentricity, nor of the libration amplitude.  As a function of eccentricity, the libration period decreases as eccentricity increases in the range 0.1 to 0.7, but the trend reverses for higher eccentricities. The shortest libration periods occur for librating orbits with $\phi_{\rm min}$ close to $\phi_0$; the libration period increases very slowly with increasing $|\phi_0-\phi_{\rm{min}}|$ when this amplitude is small; indeed, no variation is measurable for $|\phi_0-\phi_{\rm{min}}|<10^\circ$.  At higher $|\phi_0-\phi_{\rm{min}}|$, the libration period rises sharply and presumably diverges at the boundary between the symmetric and asymmetric families of librating orbits.  In the zone of the symmetric librators, the libration period drops sharply as we move away from its inner boundary and towards larger amplitude librations. 
In the region of the sharp spike of the libration period, the numerical solutions show that the test particle orbit librates very slowly in the narrow boundary region which connects the two wings of the asymmetric librators, leading to the quite long libration periods. We conjecture that this is one of the likely origins of ``resonance sticking" behavior of SDOs.

In the second resonance zone, the libration periods are shorter than in the first resonance zone (for similar values of eccentricity), and they decrease slightly with increasing libration amplitude.  

\subsection{Patterns in the $N$:1 sequence of exterior MMRs}\label{subsec:PatternsNto1}

\begin{figure}
 \centering
 \includegraphics[width=90mm]{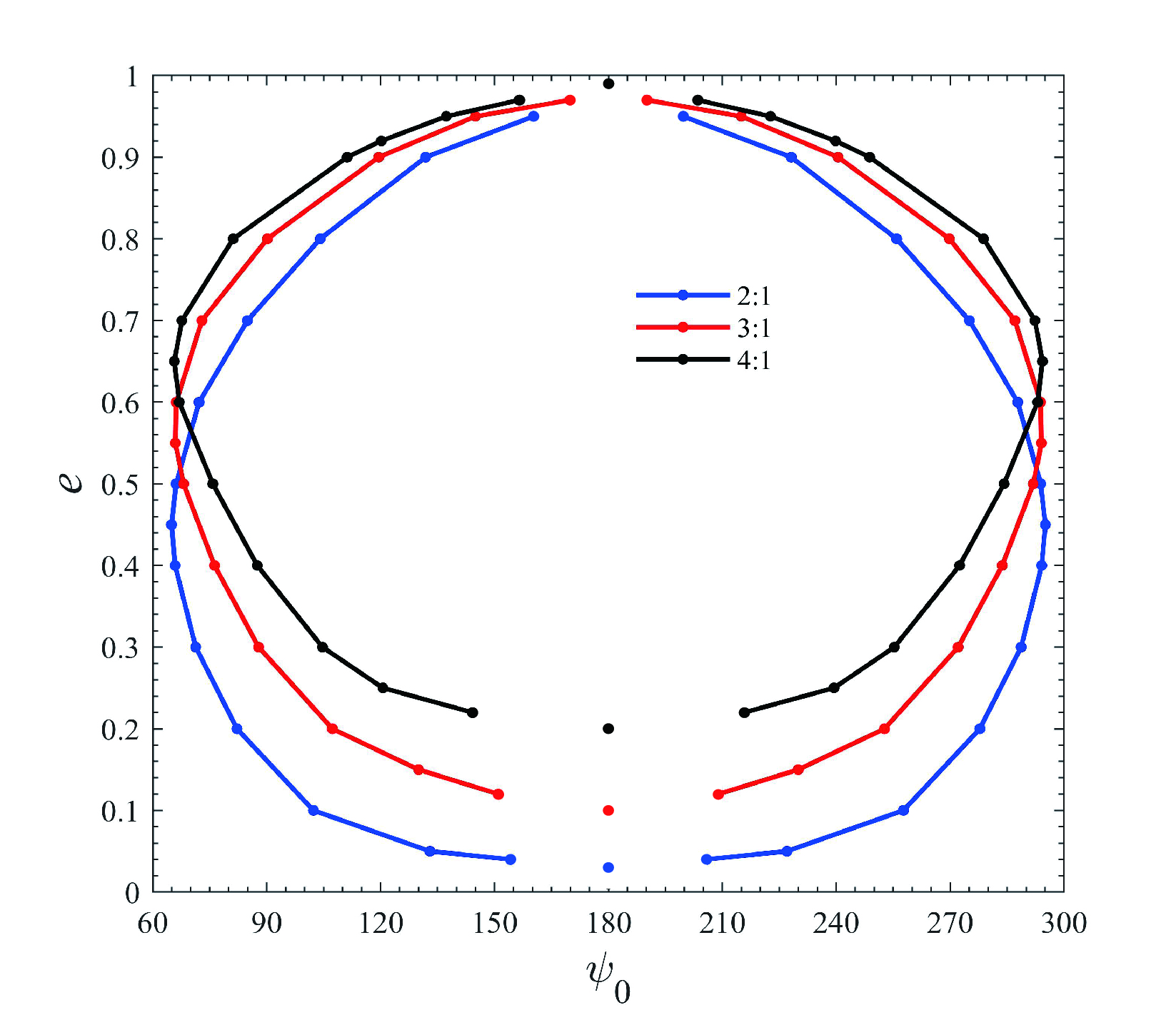}
 \caption{ The centers of the asymmetric librations in the 2:1, 3:1 and 4:1 MMRs as a function of eccentricity.}
 \label{f:cve}
\end{figure}

 For moderate-to-high eccentricities, the phase space structure of all the $N$:1 exterior resonances is qualitatively similar to that of the 2:1 resonance.
For larger $N$, the asymmetric libration zone first appears at larger values of eccentricity. The centers of the asymmetric librations as a function of particle eccentricity are shown in Figure~\ref{f:cve} for the 2:1, 3:1 and 4:1 MMRs.  Similar to the 2:1 MMR, these libration centers drift away swiftly from near $\psi_0=180^\circ$ towards an extremum near $\psi_0=\pm60^\circ$, then drift back towards $\psi_0=180^\circ$ as the eccentricity approaches unity.  
The second resonance zone first emerges at eccentricity $e_2\simeq1-0.9/a_{\rm res}$ (corresponding to perihelion distance $\sim27.1$ au), and expands in width as the eccentricity approaches unity.

\begin{figure*}
 \centering
 \includegraphics[width=110mm]{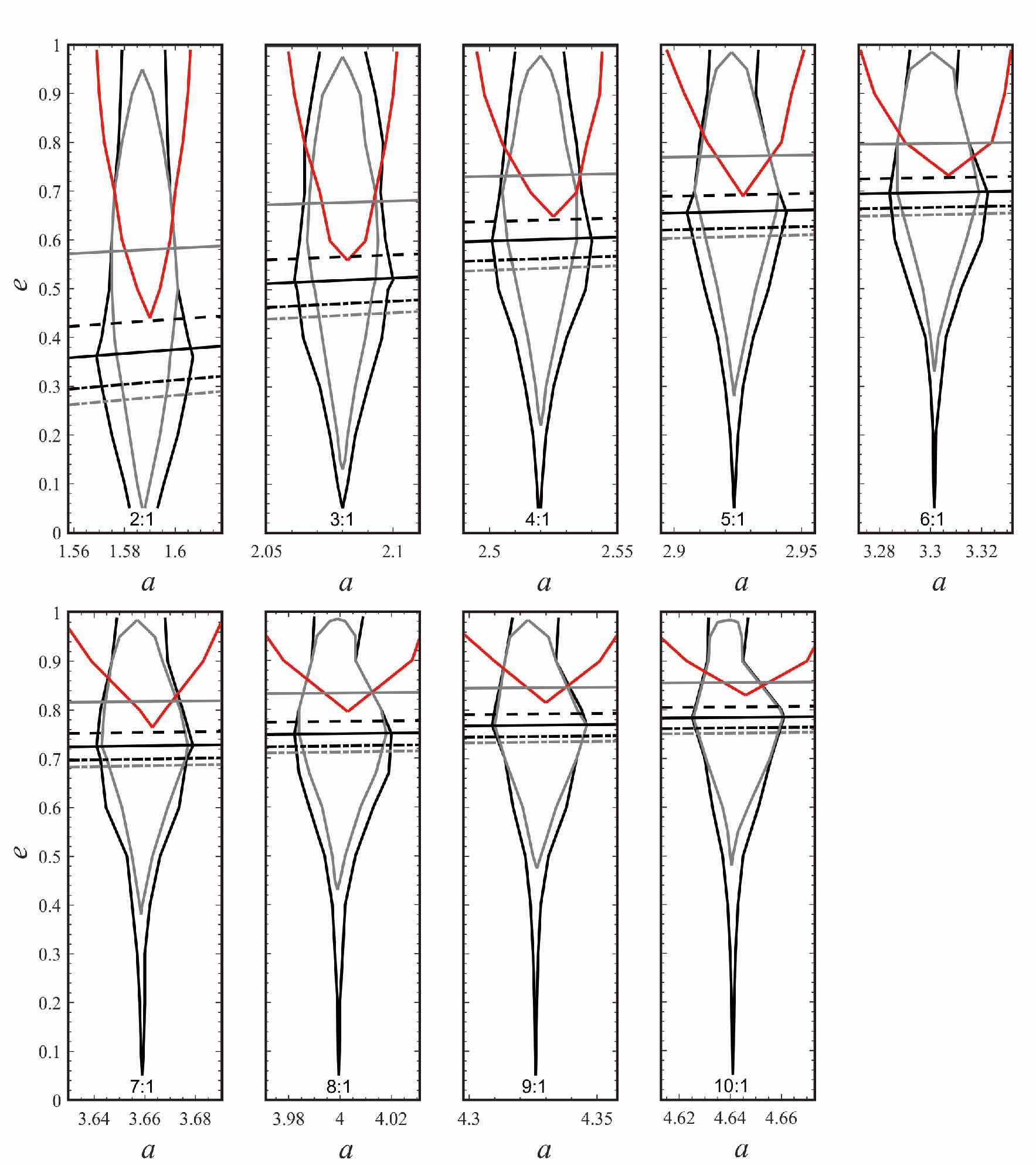}
 \caption{
 The boundaries of the stable resonance zones of the $N$:1 sequence of Neptune's exterior MMRs in the ($a,e$) plane. The area bounded by the black lines is the first resonance zone, that bounded by the grey lines is the zone of asymmetric librators within the first resonance zone, and the area bounded by the red lines is that of the second resonance zone. For reference, we also draw several curves of constant perihelion distance (see the caption for Figure~1). }
 \label{f:nt1}
\end{figure*}

\begin{figure}
\gridline{\fig{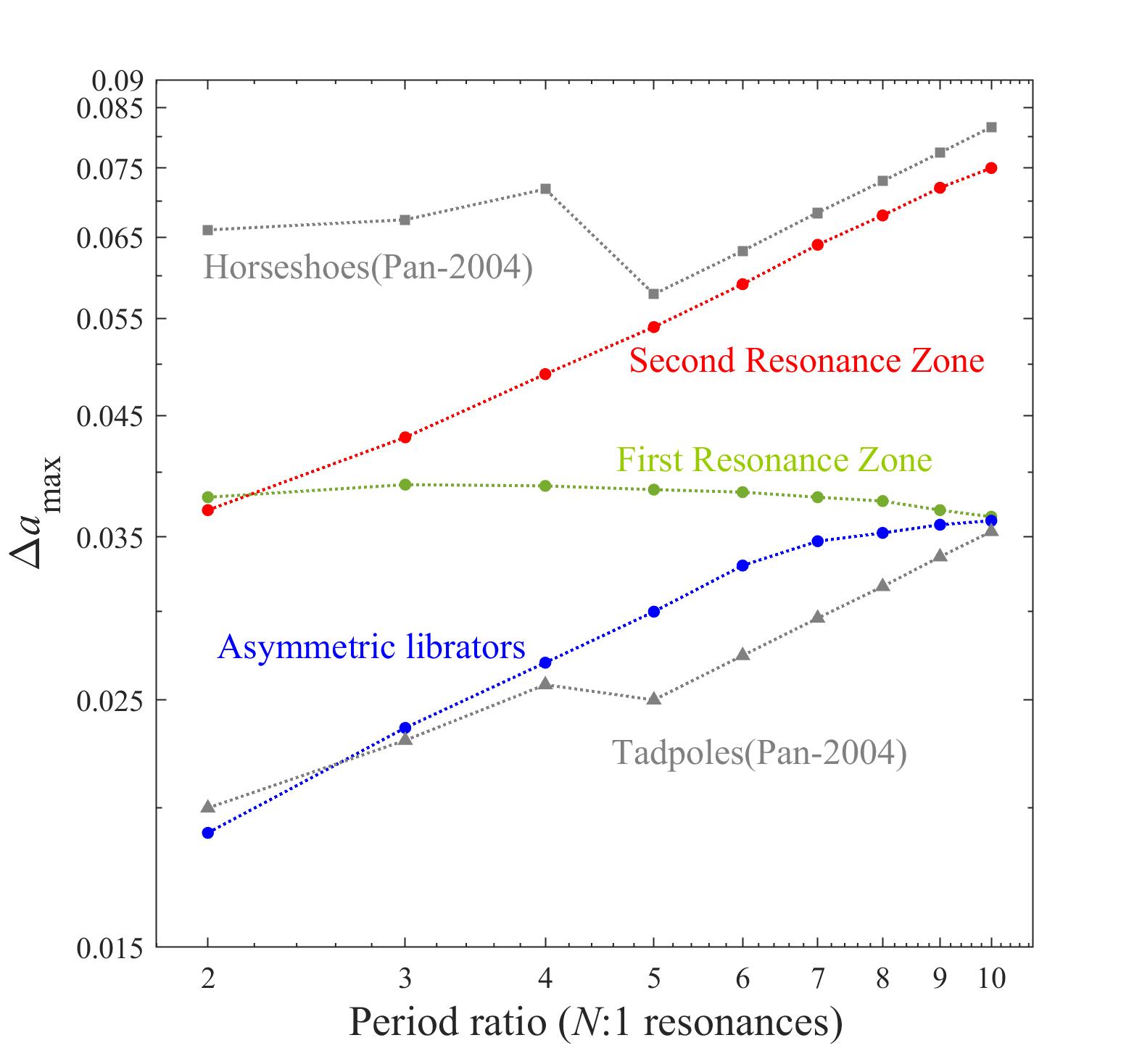}{0.5\textwidth}{(a)}
          \fig{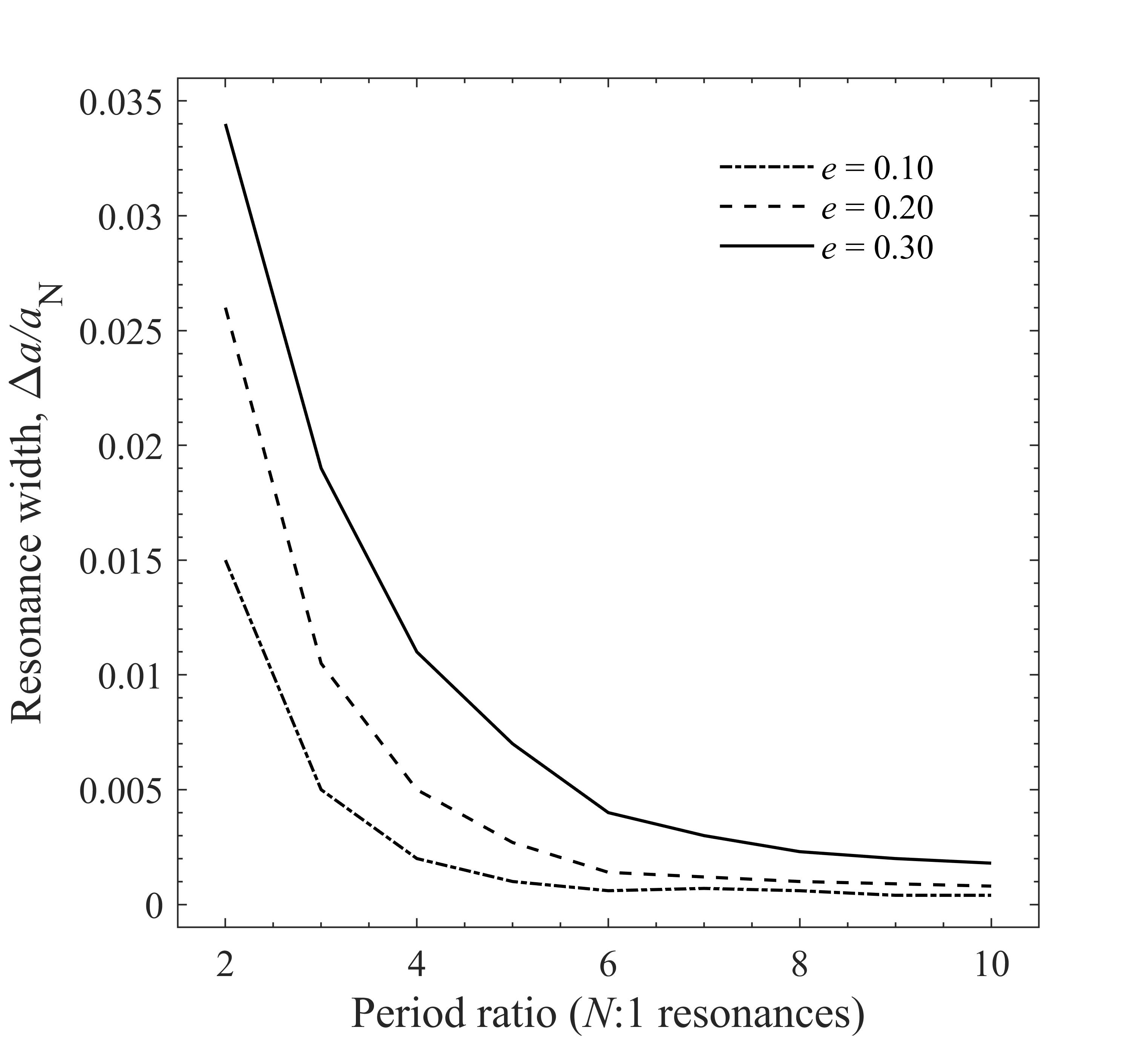}{0.5\textwidth}{(b)}
          }
\caption{
Resonance widths in semi-major axis, $\Delta a$ (in units of $a_{\rm Neptune}$) of Neptune's exterior $N$:1 resonances as a function of the particle-to-Neptune period ratio $N$. (a) The maximum width of the first resonance zone (in green), the second resonance zone (in red) and the asymmetric libration zone (in blue); also plotted (in gray) are the analytic estimates from \cite{Pan:2004} for the maximum widths of the first resonance zone (labeled ``Horseshoes") and of the asymmetric librators (labeled ``Tadpoles").
(b) The width of the first resonance zone at a few eccentricities lower than the critical planet-crossing value.
}
 \label{f:nto1width}
\end{figure}

\begin{figure}
\gridline{\fig{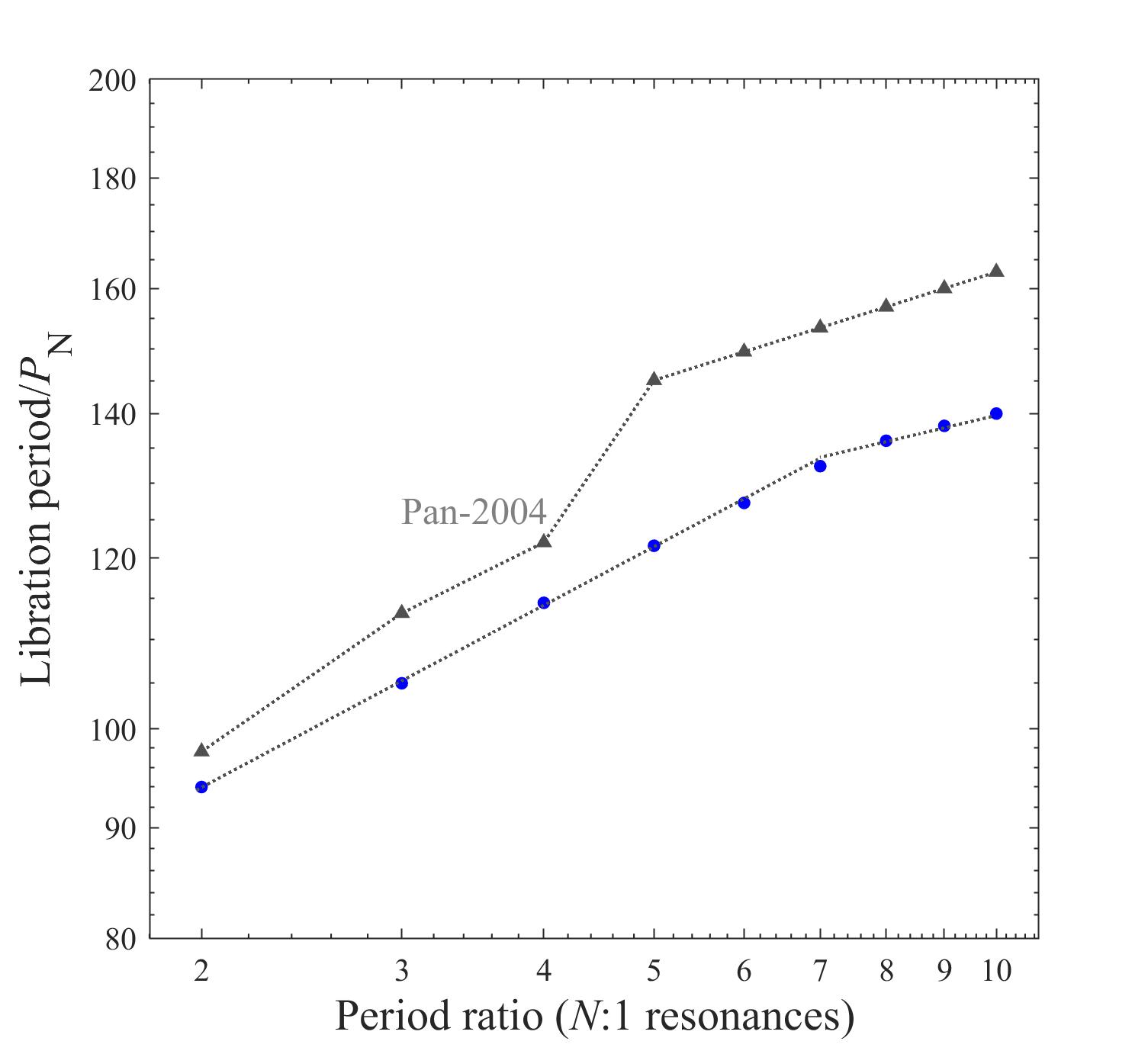}{0.5\textwidth}{(a)}
          \fig{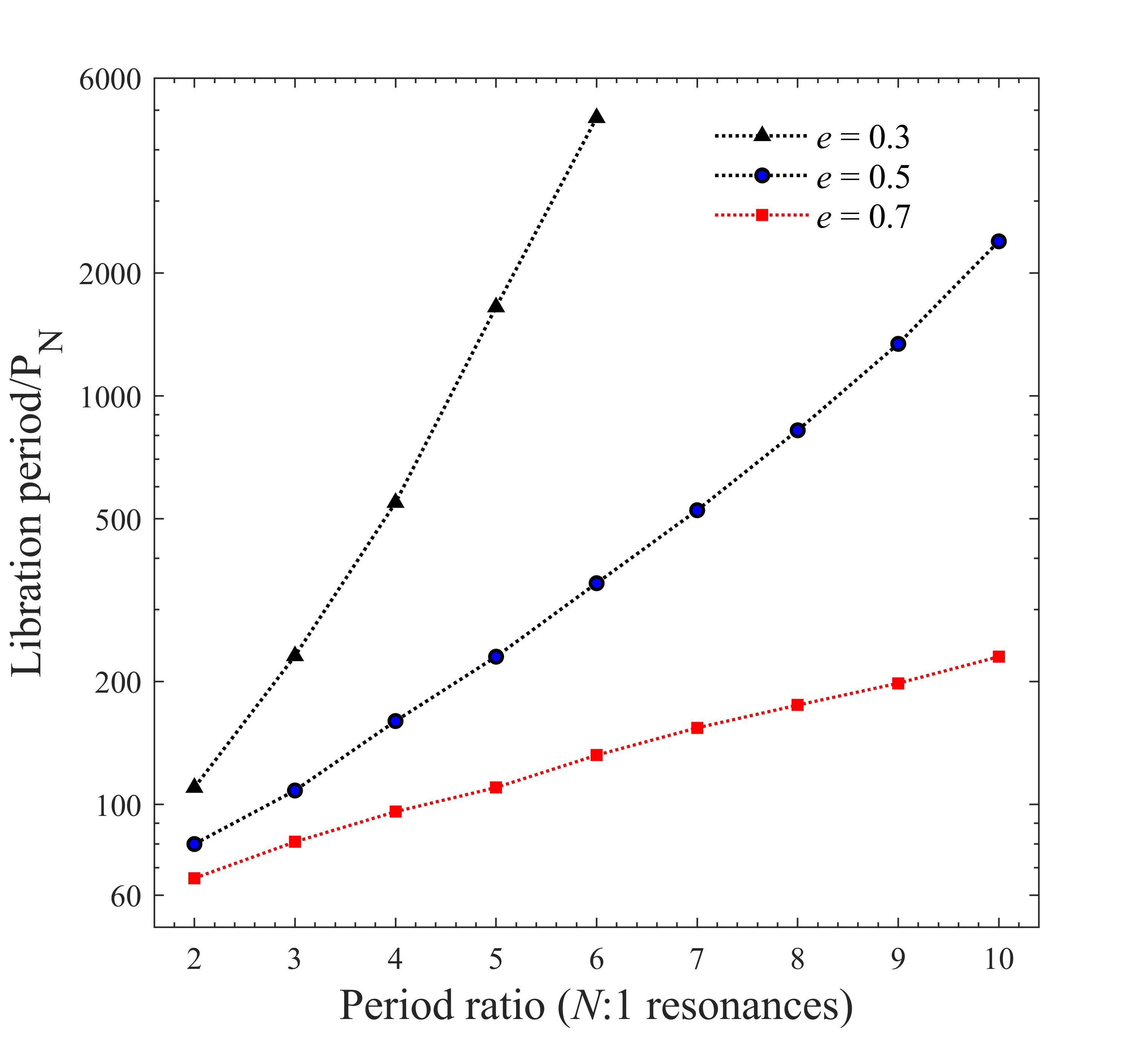}{0.5\textwidth}{(b)}
          }
\caption{
The small amplitude libration period (in units of Neptune's orbital period) of the first resonance zones of the $N$:1 sequence of Neptune's exterior resonances as a function of particle-to-Neptune period ratio, $N$. (a) At the critical eccentricities, $e_c$, the behavior is approximately a double power-law relation (blue curve); the grey curve shows the analytical estimates of \cite{Pan:2004}. (b) At eccentricities away from $e_c$ the small amplitude libration period in the first resonance zone has exponential or super-exponential dependence on $N$.
}
 \label{f:nto1LP}
\end{figure}

The boundaries in the $(a,e)$ plane of the first and second resonance zones are plotted in Figure \ref{f:nt1} for the sequence of $N$:1 resonances, $2\le N\le10$. We observe that the width, $\Delta a = a_{\rm max}-a_{\rm min}$, of the first resonance zone increases with eccentricity from $e=0.05$, reaches a maximum near $e_c$ (Eq.~\ref{e:ec}, corresponding to perihelion distance equal to Neptune's orbital radius, $\sim30.1$ au), then slowly shrinks with increasing particle eccentricity beyond $e_c$.
In Figure \ref{f:nt1} we also plot the boundaries of the asymmetric libration zone (which is a subset of the first resonance zone). 
It is apparent that the resonance widths in semi-major axis, $\Delta a$, depend on both $N$ and eccentricity.  This dependence is illustrated further in Figure \ref{f:nto1width}. 
Perhaps the most notable result is that the maximum width of the first resonance zone (which occurs near eccentricity $e_c$) is rather similar for all $N$; it changes only very slowly with increasing period ratio (see Figure~\ref{f:nto1width}a).  The 3:1 resonance has the largest width, even larger than 2:1 resonance, but all the $N$:1 resonances we examined (up to $N=10$) have quite large maximum widths. In physical units, these maximum widths are all near $\Delta a\approx1$~au. For eccentricity near $e_c$, with increasing $N$ the size of the asymmetric libration zone is an increasing fraction of the size of the first resonance zone.  

For the 3:1 resonance, the first and second resonance zones have similar maximum widths, but for the larger particle-to-Neptune period ratios, the maximum width of the second resonance zone significantly exceeds that of the first resonance zone~(Figure \ref{f:nto1width}a). In contrast with the behavior near the planet-crossing values of eccentricities, the first resonance zone's widths at lower eccentricities decrease rapidly with the increasing period ratio, as shown in Figure \ref{f:nto1width}b. For period ratios larger than 6:1 and eccentricities up to $\sim0.3$, the width of the first resonance zone does not exceed 0.1~au. 

We also observe in Figure \ref{f:nto1width}a that the maximum width of the second resonance zone increases rapidly with increasing period ratio. For the 2:1 resonance, the first and second resonance zones have similar maximum widths, but for the larger particle-to-Neptune period ratios, the maximum width of the second resonance zone significantly exceeds that of the first resonance zone. The best-fit power law, $\Delta a_{\rm{max}}=0.0136\times N^{0.49}$, provides a very good empirical approximation for the width of the second resonance zone for eccentricities near the critical Neptune-crossing value. 

The behavior of the resonance libration periods with libration amplitude and with eccentricity is qualitatively similar to that of the 2:1 MMR (Figure~\ref{f:21pe}). The small amplitude libration periods of the asymmetric librations generally increase with $N$ and decrease with increasing eccentricity, as shown in Figure~\ref{f:nto1LP}. Near the eccentricity $e_c$, the small amplitude libration period can be approximated with a double power law: $T_{\rm lib}/P_{\rm Neptune}=77.6\times N^{0.275}$, for $N$:1 resonances of $2\le N\le7$; for larger $N$, we find $T_{\rm lib}/P_{\rm Neptune}=104.1\times N^{0.130}$. At other values of eccentricity, the small amplitude libration period increases exponentially or even super-exponentially with $N$ (Figure~\ref{f:nto1LP}b).

For the special case of eccentricities close to $e_c$, we can compare our numerical results with the analytical estimates derived by \cite{Pan:2004}. The centers of libration seen in our $(\psi,a)$ Poincar\'e sections correspond to the ``generalized Lagrangian points" discussed by these authors, who also used the terminology ``generalized tadpole" and ``generalized horseshoe" orbits for the asymmetric and symmetric librators of the first resonance zone, respectively.  \cite{Pan:2004} estimated the resonance widths of tadpoles (asymmetric librators) and of horseshoes (symmetric librators) as $\Delta a_{\rm{max}}
=1.56N^{1/2}\mu^{1/2}$ 
and $\Delta a_{\rm{max}}
=3.6N^{1/2}\mu^{1/2}$, respectively, and they estimated the minimum tadpole libration period as $5.0a^{1/4}\mu^{-1/2}$ which is equivalent to $5.0N^{1/6}\mu^{-1/2}$. 
 (We have multiplied by a factor of two the resonance half-widths stated in \cite{Pan:2004}.)
In our numerical studies with the fixed value of $\mu=5.146\times10^{-5}$, the best-fit power-law relation that we found for the resonance width of the asymmetric librators (generalized tadpoles) is $\Delta a_{\rm{max}}=0.027\,N^{0.44}$; this power-law index, 0.44, is lower than the estimate of 0.5 in \cite{Pan:2004}, but overall their analytic estimate predicts widths similar to those we measured for the tadpoles; compare the curves labeled ``Asymmetric librators" and ``Tadpoles(Pan-2004)" in Figure~\ref{f:nto1width}a. For the symmetric librators (generalized horseshoes), we found the widths to be nearly independent of $N$, contrary to \cite{Pan:2004}'s estimate which predicts larger widths than we measured; compare the curves labeled ``First resonance zone" and ``Horseshoes(Pan-2004)" in Figure~\ref{f:nto1width}a.   Our numerical results on the small amplitude libration period in the first resonance zones at particle eccentricity $e_c$ are generally similar to but slightly lower than \cite{Pan:2004}'s estimates (Figure \ref{f:nto1LP}(a)).

\section{\it{N}\rm:2 resonances} \label{sec:nto2}

We computed the Poincar\'e sections for the $N$:2 sequence of Neptune's exterior MMRs, from 3:2 to 19:2, for the full range of particle eccentricities. The 3:2 MMR is the closest-in of this sequence, and its phase space structure is exemplary of all the $N$:2 sequence of MMRs.  We first describe this resonance in some detail, and then discuss the patterns of behavior of the whole sequence.

\subsection{The 3:2 MMR}\label{ss:3to2}

\begin{figure*}
 \centering
 \includegraphics[width=170mm]{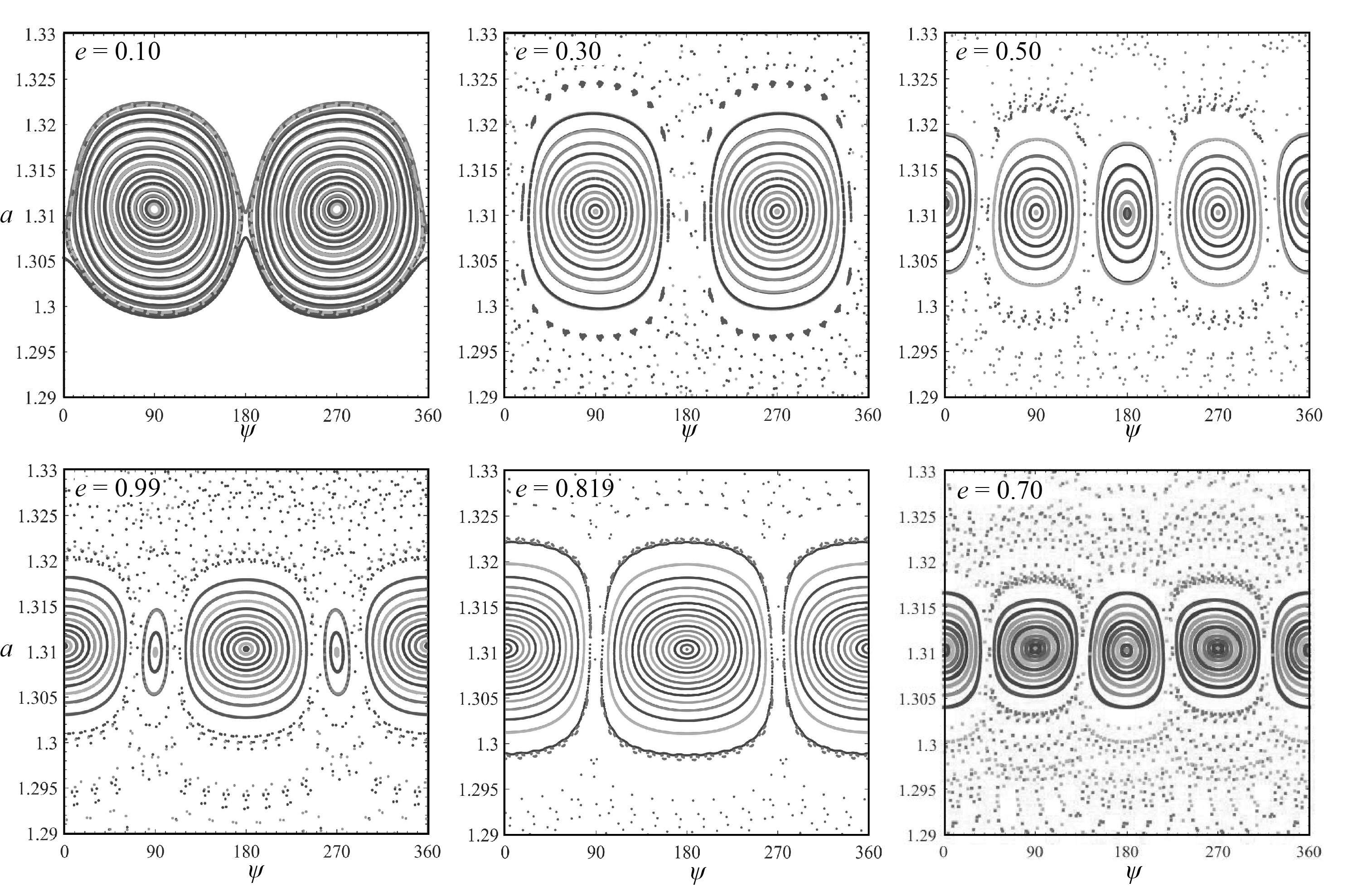}
 \caption{
The surfaces of section in ($\psi,a$) near Neptune's exterior 3:2 resonance at different particle eccentricities. Clockwise from top left: $e=0.10$; $e=0.30$; $e=0.50$; $e=0.70$; $e=0.819$; $e=0.99$. In the first two panels, only one pair of resonant islands is visible, centered at $\psi=90^\circ$ and $\psi=270^\circ$; this pair belongs to the first resonance zone. The second resonance zone first becomes visible at $e\simeq0.295$ as an additional pair of islands centered at $\psi=0$ and $\psi=180^\circ$. With increasing eccentricity above 0.295, the pair of islands of the first resonance zone shrink and vanishes at $e=0.819$, then reappears at higher eccentricity.
}
 \label{f:3t2}
\end{figure*}

\begin{figure*}
 \centering
 \includegraphics[width=160mm]{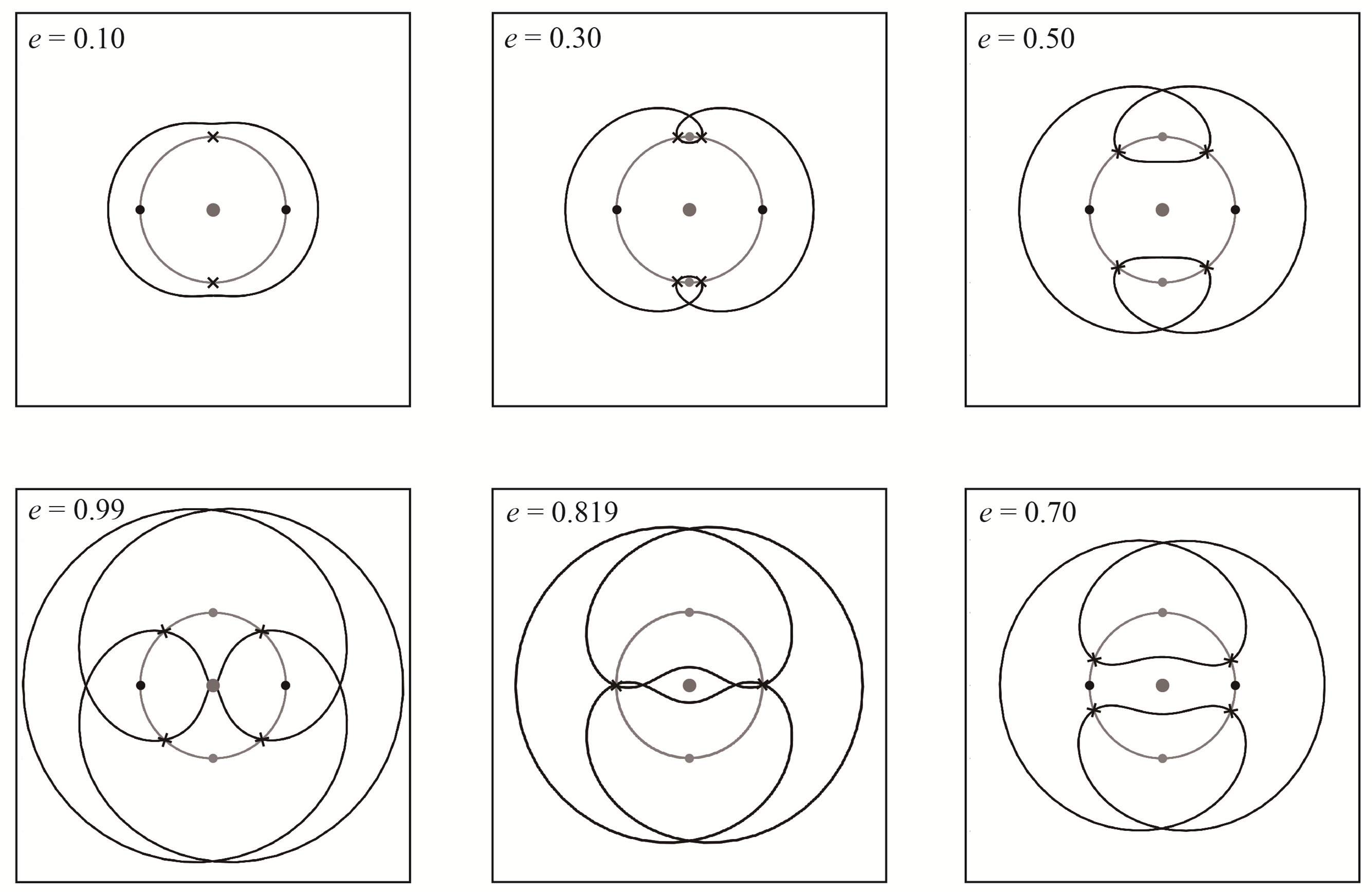}
 \caption{The geometry of the exterior 3:2 MMR in the rotating frame at different particle eccentricities. Clockwise from top left: $e=0.10$; $e=0.30$; $e=0.50$; $e=0.70$; $e=0.90$; $e=0.99$.}
 \label{f:32e}
\end{figure*}

\begin{figure}
 \centering
 \includegraphics[width=90mm]{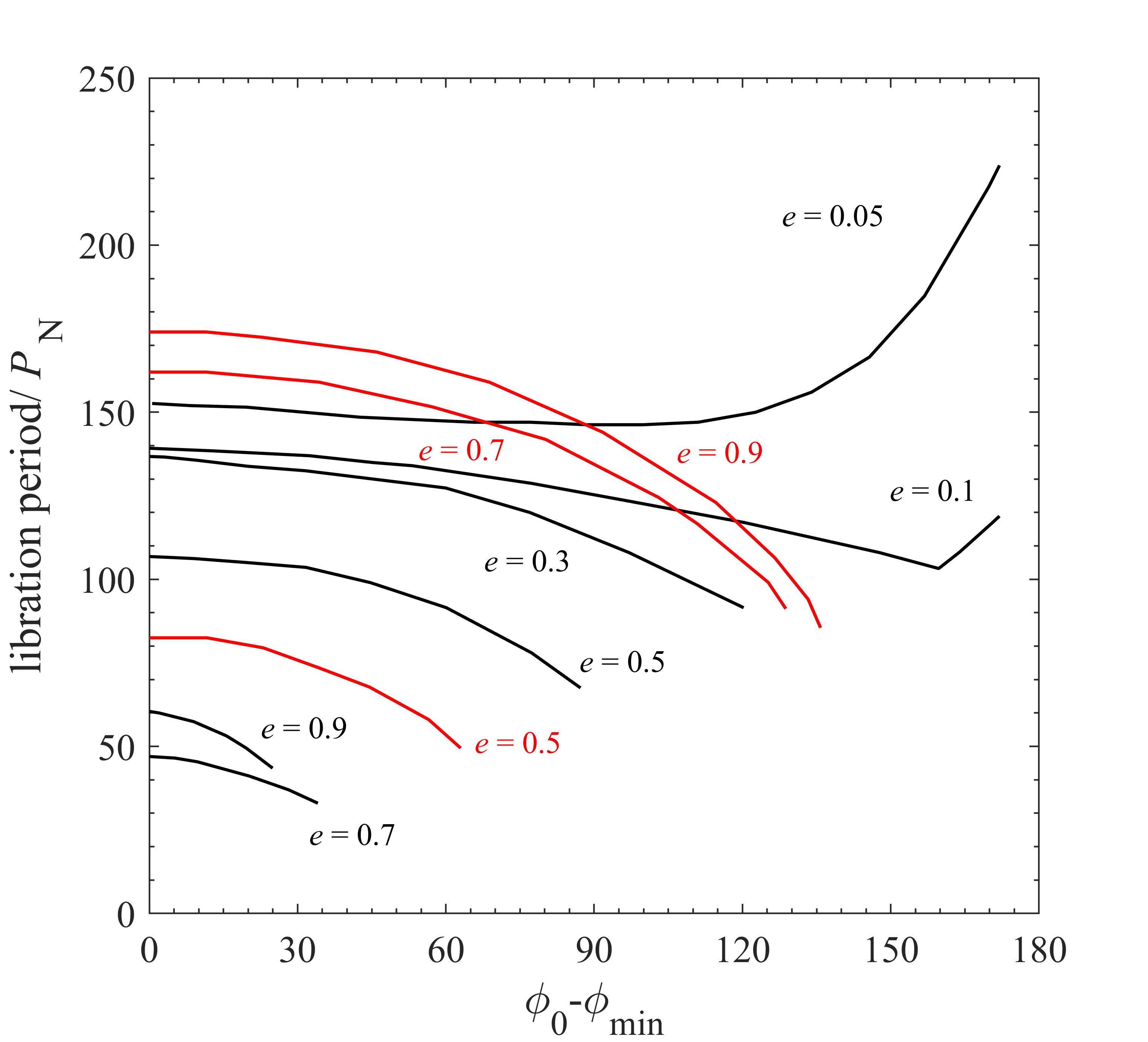}
 \caption{The libration period (in units of Neptune's orbital period) in Neptune's exterior 3:2 MMR as a function of libration amplitude of the resonant angle $\phi=2\lambda_{\rm{Neptune}}-3\lambda+\varpi$, at various particle eccentricities. Black curves indicate the first resonance zone, red curves indicate the second resonance zone.
}
 \label{f:32pe}
\end{figure}

Examining the Poincar\'e sections near the 3:2 MMR, we observe some similarities and some differences with the 2:1 MMR.  As seen in Figure \ref{f:3t2} (top left panel), at low eccentricity, there is no visible chaotic region in the Poincar\'e section in ($\psi,a$) which has a pair of stable islands composed of smooth closed curves. This is the first resonance zone. The pair of islands is centered at the two values of $\psi_0=90^\circ$ and $\psi_0=270^\circ$, but share a common central value of the semi-major axis, $a_{\rm res} = 1.309$. Unlike in the case of the 2:1 MMR, these libration centers of $\psi$ do not drift with the change of eccentricity. Also, unlike the case of the 2:1 MMR, these two islands are not separate families but form a chain of two islands, such that a single trajectory visits alternately each island in the Poincar\'e section. That is to say, the exact 3:2 resonant orbit is a ``period two" orbit. In the inertial frame, this orbit completes two circuits around the Sun in the time it closes one complete trace in the rotating frame. A pair of unstable points exists at the separatrix of this chain of two islands. These two unstable points are located at $\psi_0=0^\circ$ and $\psi_0=180^\circ$, but at slightly different values of the semi-major axis; the slight difference in the semi-major axis is due to the different positions of Neptune at the two successive perihelion passages of the test particle in this configuration. This difference gives rise to a slight difference in perihelion distance and perihelion velocity, hence a slight difference in the osculating orbital parameters at alternate perihelion passages.  

The widths of the stable islands of the 3:2 MMR increase with eccentricity, reaching a maximum at eccentricity $e_1=0.16$, where the perihelion distance is $\sim33$~au. At an eccentricity $e_2=0.295$, where the perihelion distance approaches 27 au, a new pair of stable islands appear in the surface of section (see Figure \ref{f:3t2}, top right panel).
These are centered at $\psi_0=0^\circ$ and $\psi_0=180^\circ$, in-between the old pair of stable islands observed at lower eccentricity. We call this pair the second resonance zone.  As with the old pair, this pair is also a period-two chain of islands. With increasing eccentricity, the semi-major axis widths, $\Delta a$, of these new stable islands expand at the expense of the old islands, reaching a maximum at eccentricity $e_4=0.830$. In this regime, the width, $\Delta a$, of the first resonance zone decreases with increasing eccentricity and vanishes at $e_3=0.819$ (Figure \ref{f:3t2}, bottom middle panel). At an eccentricity $e_5=0.836$, which is slightly larger than $e_4$, the stable islands of the first resonance reappear, and their sizes increase with increasing eccentricity. Meanwhile, the width of the second resonance zone shrinks slightly. Thus, when the eccentricity approaches unity, the width of the second resonance zone is only slightly larger than that of the first resonance zone (Figure \ref{f:3t2}, bottom left panel).

The transitions with increasing eccentricity in the Poincar\'e sections of Neptune's 3:2 MMR described above are correlated with the changes in the geometry of the trace of the resonant orbit in the rotating frame.  In Figure \ref{f:32e}, we show the trajectories of the particle's 3:2 resonant orbit in the rotating frame at different particle eccentricities. We observe that the first appearance of the second resonance zone occurs when the eccentricity slightly exceeds the Neptune-crossing eccentricity, and the small initial size of this zone is due to the small length of the arc of the Sun-centered unit circle which is cut by the particle's perihelion lobes.  With increasing eccentricity, the length of arc enclosed by the perihelion lobes increases, and the size of the second resonance zone increases correspondingly, at the expense of the first resonance zone.  The vanishing of the first resonance zone occurs when the two perihelion lobes touch each other, when the eccentricity approaches $e_3=0.819$.  At even higher eccentricity, the two perihelion lobes intersect each other, creating new arcs for the reappearance of the first resonance zone.

The behavior of the libration period of the resonant angle, $\phi=2\lambda_{\rm{Neptune}}-3\lambda+\varpi$, is rather complicated, as shown in Figure~\ref{f:32pe}.  In the first resonance zone, for the eccentricity range 0.1--0.7, the libration period decreases with increasing eccentricity and also decreases with increasing libration amplitude, $|\phi_0-\phi_{\rm{min}}|$; 
however, the librating orbits of $e=0.90$ have longer libration periods than those of $e=0.70$.
In the second resonance zone, the libration period decreases with increasing libration amplitude, $|\phi_0-\phi_{\rm{min}}|$ (similar to the behavior in the first resonance zone), but increases with increasing eccentricity (unlike in the first resonance zone). 

 \cite{Gallardo:1998} numerically computed the libration periods in Neptune's 3:2 MMR for eccentricities up to 0.35; our results for that range of eccentricities are consistent with their findings.

\subsection{Patterns in the $N$:2 sequence of exterior resonances} \label{sec:nto2patterns}

\begin{figure*}
 \centering
 \includegraphics[width=110mm]{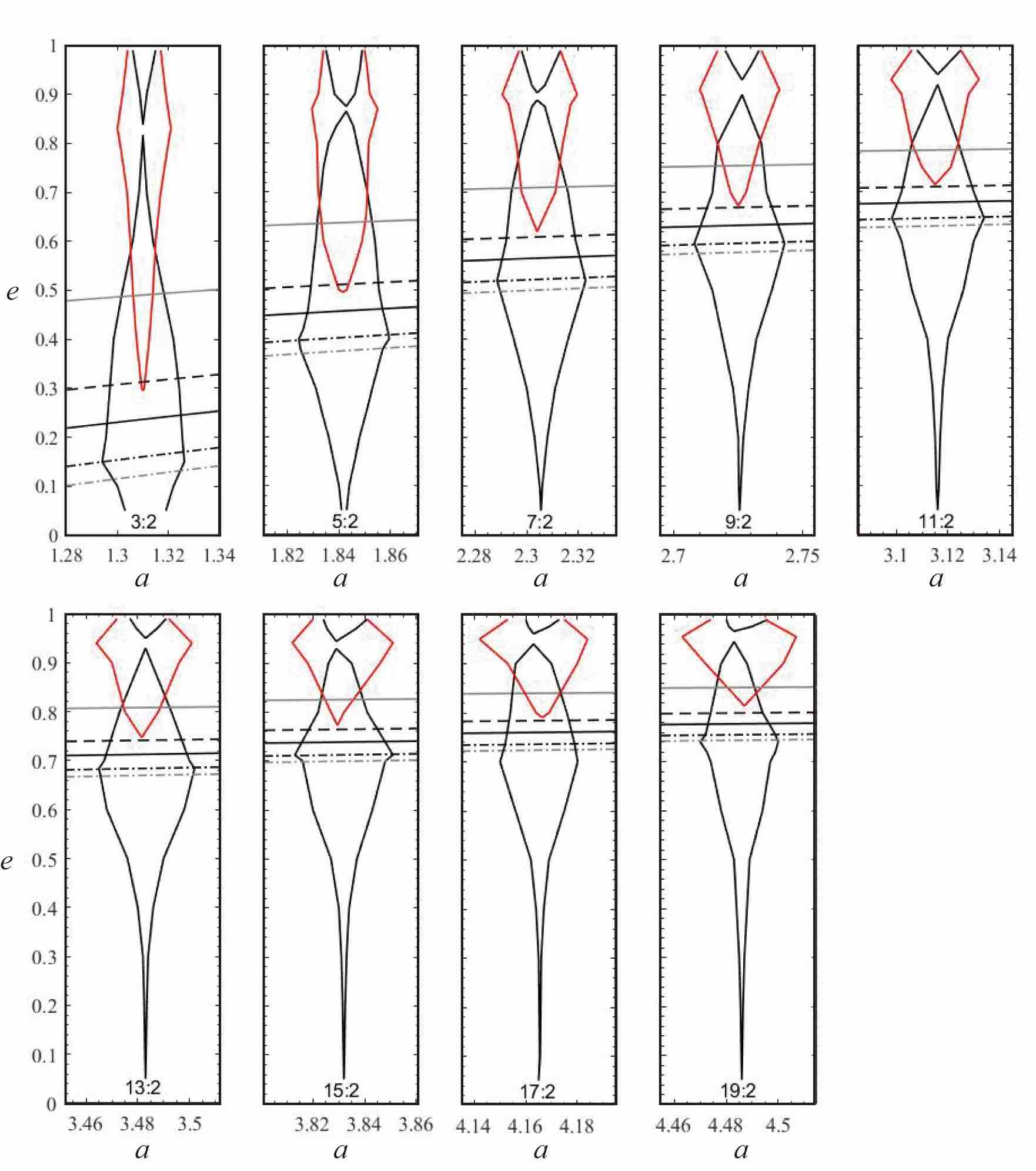}
 \caption{The boundaries of the stable resonance zones of the $N$:2 sequence of Neptune's exterior resonances of  different period ratios in the ($a,e$) plane. For reference, we also draw several curves of constant perihelion distance (see the caption for Figure~1).}
 \label{f:nt2}
\end{figure*}

\begin{figure}
\gridline{\fig{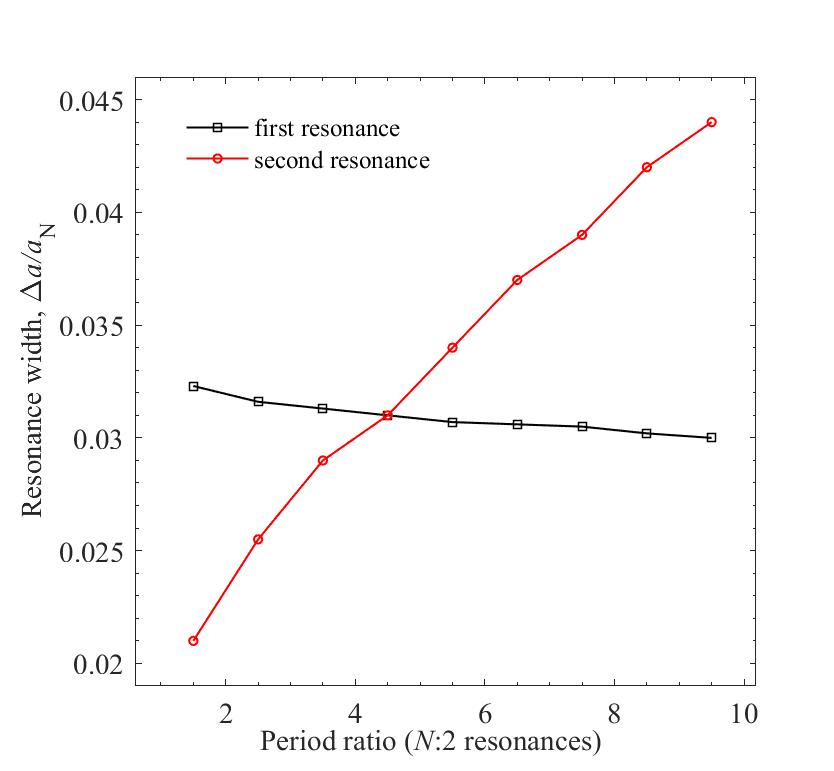}{0.5\textwidth}{(a)}
          \fig{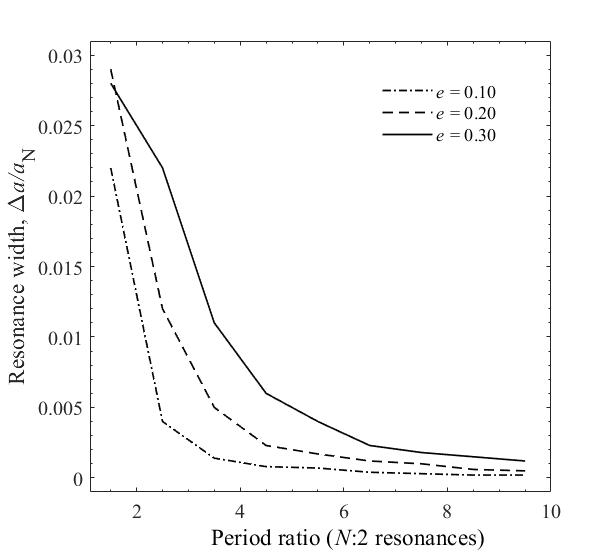}{0.5\textwidth}{(b)}
          }
\caption{
The widths of the first resonance zones of the $N$:2 sequence of Neptune's exterior resonances. (a) The maximum widths of the first resonance zone (which occurs at eccentricity $e_1\simeq 1-1.1a_{\rm Neptune}/a_{\rm res}$, in black) and the second resonance zone (which occurs at eccentricity $e_4$, in red). (b) The widths of the first resonance zones at a few other eccentricity values.
}
 \label{f:wi2}
\end{figure}

By examining the Poincar\'e sections for the full range of eccentricities for the $N$:2 sequence of Neptune's exterior MMRs, 3:2 to 19:2, we ascertained the boundaries of the stable libration zones.  These are shown in Figure \ref{f:nt2} in the ($a,e$) plane. Similar to the 3:2 resonance, every $N$:2 resonance has a first resonance zone and a second resonance zone, each with a pair of stable islands. With increasing values of the test particle eccentricity, the resonance structure in the Poincar\'e sections show five transitions at eccentricity values $e_1,e_2,...,e_5$ which are similar to those described above for the 3:2 MMR. The first resonance zone consists of a pair of stable islands centered at $\psi_0=90^\circ$ and $\psi_0=270^\circ$.  Starting at a low eccentricity, we observe that its width, $\Delta a$, increases with increasing value of eccentricity, its largest extent occurs at an eccentricity, $e_1\simeq1-1.1/a_{\rm res}$ (corresponding to perihelion distance $\sim33.1$ au). The stable islands then shrink and finally vanish at an eccentricity $e_3$.  Meanwhile, at an eccentricity, $e_2\simeq1-0.9/a_{\rm res}<e_3$ (corresponding to perihelion distance $\sim27.1$ au), the second resonance zone emerges, consisting of a new pair of stable islands centered at $\psi=0^\circ$ and $\psi=180^\circ$, located in-between the stable islands of the first resonance. The second resonance zone expands with increasing eccentricity from $e_2$, reaches a maximum size at eccentricity $e_4(>e_3)$, and then shrinks with increasing eccentricity. At an eccentricity, $e_5(>e_4)$, the pair of stable islands of the first resonance zone (centered at their original locations, $\psi_0=90^\circ$ and $\psi_0=270^\circ$) emerges again. These expand in width, while the islands of the second resonance zone shrink as the eccentricity approaches unity.  

A detail that we note in Figure \ref{f:nt2} is that for the more distant MMRs the boundaries of the resonance zones at higher eccentricity are noticeably not symmetric about $a_{\rm{res}}$. Similar asymmetries are also visible in Figure \ref{f:nt1} for the distant $N$:1 resonances. These asymmetries are also noticeable in the corresponding Poincar\'e sections where we can see that the lower and upper half of the stable resonant islands are not exactly symmetric. These asymmetries arise because librating trajectories are not symmetric about the Sun-Neptune line. At one extreme of the libration zone, the test particle's closest approach to Neptune occurs somewhat before perihelion whereas at the other extreme of the libration zone the closest approach occurs somewhat after perihelion. (These two extremes  can be visualized by considering the geometry in the case when Neptune is at a location close to the unstable point in the upper right and the upper left in Figure~\ref{f:32e}, $e=0.5$.) Consequently the osculating orbital elements librate asymmetrically about the stable periodic orbit at the center of a resonant island. The differences are small, but more pronounced for the more distant MMRs.

The variation of the resonance width with particle-to-Neptune period ratio is illustrated in
Figure \ref{f:wi2}.  We plot the maximum widths $\Delta a$ (which occur at $e_1\simeq1-1.1/a_{\rm res}$, perihelion distance $\sim33.1$~au) of the first and second resonance zones in Figure \ref{f:wi2}a. Notably, the maximum width of the first resonance zone is almost the same for all $N$:2 resonances; it decreases very slowly with increasing particle-to-Neptune period ratio. Even more surprisingly, the maximum width of the second resonance zone increases markedly with increasing particle-to-Neptune period ratio. For the 9:2 resonance, the first and the second resonance zones have almost the same maximum widths. For the higher period ratios, the maximum width of the second resonance zone even exceeds that of the first resonance zone;  this maximum is in the regime of Uranus-crossing eccentricities. 

In Figure \ref{f:wi2}b, we plot the $N$:2 resonance widths at lower and moderate eccentricities, $e=0.10, e=0.20$ and $e=0.30$. In this eccentricity regime, the resonance widths drop quickly with increasing particle-to-Neptune period ratio; only those $N$:2 resonances with period ratio below about 6 have widths exceeding 0.1 au.

\begin{figure}
\gridline{\fig{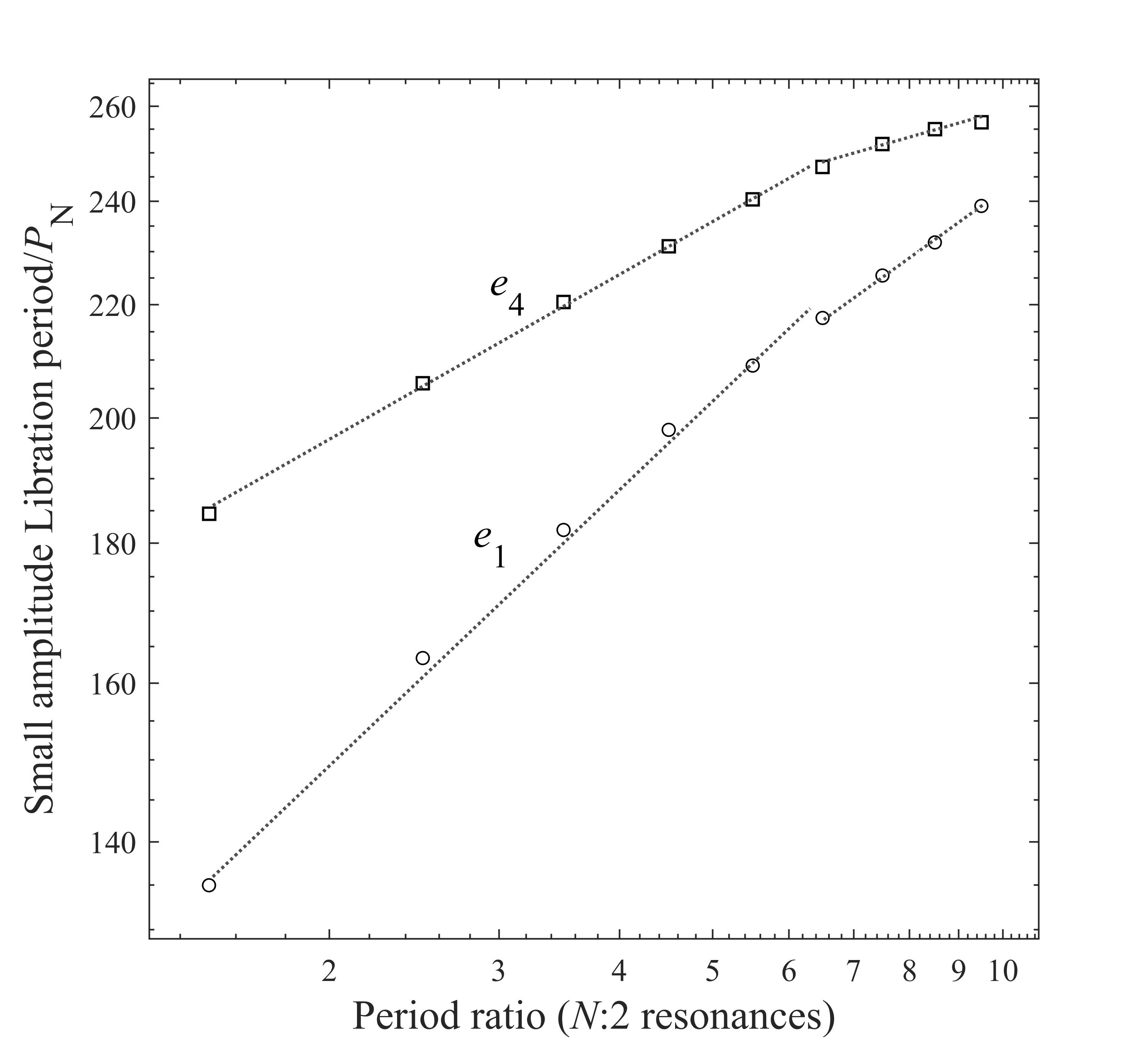}{0.5\textwidth}{(a)}
          \fig{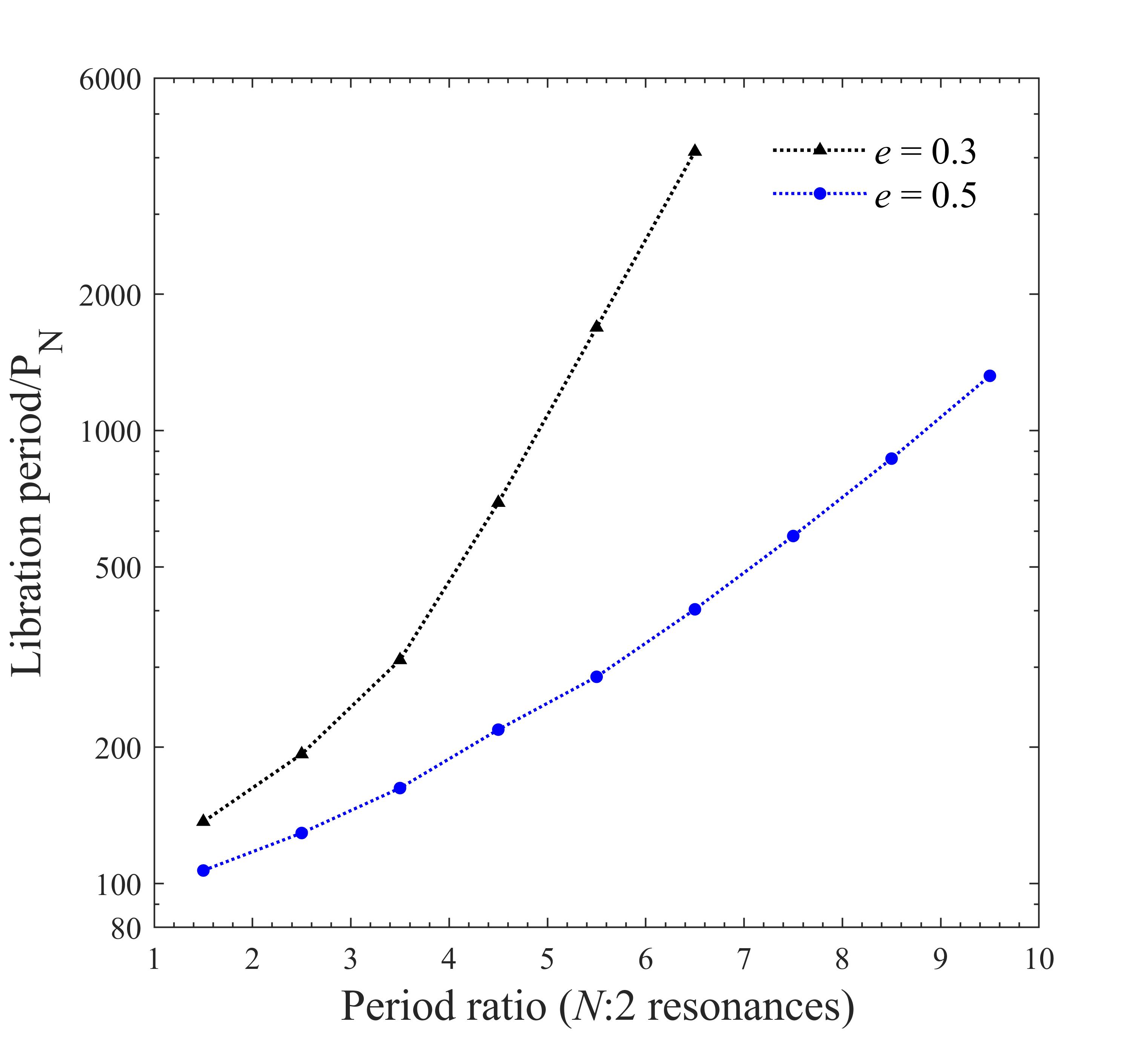}{0.5\textwidth}{(b)}
          }
\caption{
The small amplitude libration period (in units of Neptune's orbital period) of the first resonance zones of the $N$:2 sequence of Neptune's exterior resonances as a function of particle-to-Neptune period ratio, $N/2$. (a) At the critical eccentricities, $e_1$ and $e_4$, the behavior is approximately a double power-law. (b) At eccentricities away from $e_1$ the small amplitude libration period in the first resonance zone has exponential or super-exponential dependence on $N/2$.
}
 \label{f:nto2LP}
\end{figure}

The variation of the libration period with libration amplitude of the $N$:2 resonances is qualitatively similar to that of the 3:2 MMR (Figure~\ref{f:32pe}), but with an overall scale factor that can be deduced from the small amplitude libration periods, $|\phi_0-\phi_{\rm{min}}|<10^\circ$; the latter are plotted in Figure~\ref{f:nto2LP} as a function of the particle-to-Neptune period ratio, for several values of eccentricity.
We observe that, in the first resonance zone, the small amplitude libration period, $T_{\rm lib}$, at eccentricity $e_1$ (where the width of the first resonance zone reaches a maximum) approximately follows a double power-law relation with particle-to-Neptune period ratio: $T_{\rm lib}=118.3P_{\rm Neptune}\times l^{0.335}$ for $l\le 5.5$ 
and $T_{\rm lib}=135.2 P_{\rm Neptune}\times l^{0.253}$ for $l\ge 6.5$ ($l$ denotes period ratio, $l=N/2$). 
However, for eccentricity away from $e_1$ the small amplitude libration period has an approximately exponential or even super-exponential dependence on the particle-to-Neptune period ratio (Figure~\ref{f:nto2LP}b); the best-fit exponential functions for $l\ge3.5$ are found to be $T_{\rm lib}/P_{\rm Neptune}=33.99\times2.60^l+98.00$ for $e=0.3$, and $T_{\rm lib}/P_{\rm Neptune}=35.84\times1.55^l+104.00$ for $e=0.50$.  For the second resonance zone, the small amplitude libration period at $e_4$ (where the width of the second resonance reaches a maximum) also is approximately a double power-law relation with particle-to-Neptune period ratio: $T_{\rm lib}/P_{\rm Neptune}=171.0\times l^{0.200}$ for $l\le 5.5$
and $T_{\rm lib}=200.3 P_{\rm Neptune}\times l^{0.108}$ for $l\ge 6.5$.

\section{Summary and Discussion} \label{sec:sd} 

We investigated the phase space structure of many of Neptune's exterior MMRs over nearly the entire range of eccentricities, using non-perturbative numerical analysis and visualizations with Poincar\'e sections of the circular planar restricted three body model of the Sun, Neptune and test particle. We presented the Poincar\'e sections in the $(\psi,a)$ plane, where $\psi$ is the longitude separation of the planet from the test particle and $a$ is the test particle's osculating semi-major axis at the particle's perihelion passage; this allows easy identification of the resonance libration zones and measurement of their widths in semi-major axis. Our investigation spanned all the $N$:1 and $N$:2 sequence of resonances in the semi-major axis range 33--140 au (particle-Neptune period ratios up to $\sim$~10), and a few $N$:3, $N$:4 and $N$:5 resonances.
We find that, in general, there are two distinct resonance zones in phase space. The first resonance zone exists over almost the entire range of eccentricities but the second resonance zone exists only in the high eccentricity regime for eccentricities exceeding the planet-crossing value. The first and second resonance zones differ from each other in that their libration centers are shifted in phase. In the first resonance zone, conjunctions with the planet are stable near the aphelion of the particle, whereas in the second resonance zone the conjunctions are stable near the particle's perihelion. In terms of the usual resonant angle, Eq.~\ref{e:phi}, the first resonance zone librations are centered at $\phi=180^\circ$, whereas the second resonance zone librations are centered at $\phi=0$. (For the $N$:1 resonances, the first resonance zone is further split into symmetric and asymmetric librators; see below.) In the high eccentricity planet-crossing regime, a particle in the first resonance zone avoids close encounters with the planet by reaching perihelion at true longitudes well separated from the planet's true longitude. This is the well known ``phase protection" mechanism for many resonant populations in the Kuiper belt, such as the Plutinos and Twotinos~\citep{Malhotra:1995,Malhotra:1996}. In the second resonance zone, the particle avoids encounters with the planet differently: by ``looping around the planet" during the part of its orbit that is interior to the planet's; this physical explanation is illustrated by the trace of the high eccentricity resonant orbit in the rotating frame (for examples, see Figures~\ref{f:21e} and \ref{f:32e}).  

For the first resonance zone, the width in semi-major axis, $\Delta a$, is quite small at low eccentricity, rises sharply with increasing eccentricity, reaches a maximum near or slightly below the critical planet-crossing eccentricity $e_c$ (Eq.~\ref{e:ec}), then decreases again; for closer-in resonances, it vanishes and then reappears again at higher eccentricities whereas for the more distant resonances its width monotonically decreases as eccentricity approaches unity (Figures \ref{f:f1b}, \ref{f:nt1}, \ref{f:nt2}). Previous studies based on perturbative approaches with a truncated or numerically averaged disturbing potential to isolate individual resonances \citep[e.g.][]{Murray:1999SSD, Morbidelli:1995} have also found similar trends of resonance width with eccentricity, but with two significant differences: (i) the resonance width was found to diverge at the critical eccentricity, $e_c$, and (ii) for $e>e_c$, the non-monotonic variation of resonance width of the closer-in resonances was not observed because resonance widths were not computed for eccentricities exceeding the Uranus-approaching values. (\citet[2019]{Gallardo:2006b} reported non-monotonic variation of the numerically computed ``resonance strength" for some resonances at high eccentricities.) Our non-perturbative method finds finite, non-divergent, widths near $e_c$ because it naturally accounts for the interaction of neighboring resonances; these interactions quench the singularity at $e_c$ which occurs in the numerically averaged disturbing potential. The non-monotonic variation of the resonance width is owed to the geometry of the resonant orbit in the rotating frame and the existence of the second resonance zone; this zone has not been discussed in previous studies.

The second resonance zone exists at eccentricities exceeding the critical planet-crossing value. For distant resonances, its width  $\Delta a$ increases monotonically with increasing eccentricity, but for closer-in resonances, its width reaches a maximum then decreases and even vanishes and then reappears at higher eccentricities. The second resonance zone co-exists with but competes for phase space volume with the first resonance zone. We observe that the various transitions in the phase space structure, including the changes in the resonance widths with eccentricity, are related to the geometrical properties of the trace of the resonant orbit in the rotating frame. These results -- particularly the characteristics of the second resonance zone and its influence on the first resonance zone -- shed new light on Neptune's exterior resonances, beyond what has been previously studied with either perturbative analytical theory in the low eccentricity regime~\citep[e.g.][]{Murray:1999SSD} or with numerical methods to estimate resonance strengths and widths~\citep[e.g.][]{Morbidelli:1995,Robutel:2001,Gallardo:2018}.   

The small amplitude libration periods are on the order of $\sim10^2$ times Neptune's orbital period, shorter in the second resonance zone than in the first resonance zone. The small amplitude libration period in the first resonance zone increases monotonically with $N$; it has power-law dependence on $N$ for the planet-crossing minimum eccentricity $e_c$, but has exponential or even super-exponential dependence on $N$ for eccentricities away from $e_c$ (Figure~\ref{f:nto1LP} and Figure~\ref{f:nto2LP}). 
The dependence of the libration period on libration amplitude and eccentricity is rather complex (Figure~\ref{f:21pe} and Figure~\ref{f:32pe}). These trends are notably in contrast with the behavior of the common pendulum model for resonance (see, e.g., \cite{Murray:1999SSD}) and also in contrast with the behavior of the ``second fundamental model for resonance''~\citep{Henrard:1983}; in both of these models, the libration period monotonically increases with increasing libration amplitude, indicating that these models are oversimplified for the cases studied here. 

We can compare in more detail the resonance boundaries in $(a,e)$ plotted in our Figures~\ref{f:f1b}, \ref{f:nt1} and \ref{f:nt2} with those found by \citet[their Figure 1]{Morbidelli:1995}. One major difference is that \cite{Morbidelli:1995} did not consider the existence of the second resonance. For the first resonance zone, our results for the trends with eccentricity are similar but we find significantly narrower maximum widths. For example, the maximum width of the 3:1 MMR in \cite{Morbidelli:1995} is approximately 3 AU, but in our Figure~\ref{f:nt1} it is about 1.2 AU. The reason for this difference is that we measured the extent of the stable resonance island whereas \cite{Morbidelli:1995} did not account for the unstable chaotic region around the stable island.  Another significant difference is the eccentricity where the maximum width of an MMR occurs. \cite{Morbidelli:1995} find that all of Neptune's MMRs have maximum width at Neptune-crossing eccentricity $e=1-a_N/a_{\rm res}$, whereas we find that only the $N$:1 resonances have maximum width at Neptune-crossing eccentricity, but the $N$:2 and $N$:3 resonances have maximum widths at progressively lower eccentricity, at $e=1-1.1a_N/a_{\rm res}$ and $e=1-1.15a_N/a_{\rm res}$, respectively.

 It is to be noted that in the real solar system, much of the second resonance zone is interior to Uranus' orbit and is rendered unstable due to the effects of Uranus and other interior planets not considered in our simple model. Nevertheless, an important minor sector of the second resonance zone lies at lower eccentricities where the perihelion distance may be large enough for long-term stability. At the larger, Uranus-crossing eccentricities, the second resonance zone contributes to temporary resonance sticking. An example of this was reported for the minor planet 2013 UR15 \citep{Malhotra:2018}.

\subsection{Implications for SDOs}\label{subsec:imps}

\subsubsection{The second resonance zone seriously affects the dynamics of SDOs}\label{subsubsec:zone2}
At planet-crossing eccentricities, the existence of the second resonance zone affects the dynamics of the SDOs in two ways. Firstly, it squeezes the angular extent of the first resonance zone such that $\psi$ (as well as the resonant angle $\phi$) is limited to values well away from 0$^\circ$ (see Figures~\ref{f:2t1} and \ref{f:3t2}); this has the effect of pushing the resonance boundary far away from the location of the planet. From this result, we learn that the main reason that resonance sticking increases the dynamical lifetimes of SDOs is that the angular extent of the first resonance zone is small enough that resonant SDOs are effectively excluded from close approaches with Neptune. (Only the action of long term chaotic diffusion away from the resonance boundaries, possibly with significant influence of the collective effects of all the planets, allows eventual destabilizing close approaches to Neptune.)
Secondly, it offers a new mechanism for planet-crossing orbits to avoid destabilizing close encounters with Neptune by ``looping around the planet" during the portion of the orbit interior to the planet's orbit; this is distinct from the previously understood ``phase protection" mechanism in the first resonance zone which maintains the perihelion of the particle away from the longitude of the planet. This creates additional sticking opportunities for SDOs, in addition to those associated with the previously recognized first resonance zone.  

\begin{table}
\caption{
The area, in the $(a,e)$ plane, of the first resonance zone of Neptune's exterior MMRs.
}

\begin{center}
\begin{tabular}{ccccccc}
\hline
 MMR & $a_{\rm{res}}/a_{\rm{N}}$ & $a_{\rm{res}}/\rm{au}$ & area (au) & cumulative\, area (au) & area (au) & cumulative\,area (au)\\
          &                                         &   & (up to $q>26$~au) & & ($26$ au $<q<38$ au) & \\
\hline

6:5     &   1.129    &33.99  & 0.0485  &  0.0485 & 0.0485 & 0.0485 \\
5:4     &   1.160    &34.93  & 0.0712  &  0.1197 & 0.0712 & 0.1197 \\
4:3     &   1.211    &36.46  & 0.1430  &  0.2628 & 0.1430 & 0.2628 \\
7:5     &   1.251    &37.67  & 0.0865  &  0.3492 & 0.0865 & 0.3492 \\
3:2     &   1.310    &39.44  & 0.3328  &  0.6821 & 0.3328 & 0.6821 \\
8:5     &   1.368    &41.18  & 0.1044  &  0.7864 & 0.1013 & 0.7833 \\
5:3     &   1.406    &42.31  & 0.1626  &  0.9491 & 0.1618 & 0.9451 \\
7:4     &   1.452    &43.71  & 0.1228  &  1.0719 & 0.1098 & 1.0549 \\
2:1     &   1.587    &47.78  & 0.3346  &  1.4065 & 0.2474 & 1.3023 \\
7:3     &   1.759    &52.95  & 0.1762  &  1.5827 & 0.1294 & 1.4317 \\
5:2     &   1.842    &55.44  & 0.2797  &  1.8624 & 0.1897 & 1.6215 \\
8:3     &   1.923    &57.88  & 0.1900  &  2.0524 & 0.1305 & 1.7520 \\
3:1     &   2.080    &62.61  & 0.3408  &  2.3932 & 0.2027 & 1.9546 \\
7:2     &   2.305    &69.39  & 0.2669  &  2.6601 & 0.1488 & 2.1034 \\
4:1     &   2.520    &75.85  & 0.3325  &  2.9925 & 0.1760 & 2.2794 \\
9:2     &   2.726    &82.04  & 0.2484  &  3.2409 & 0.1223 & 2.4017 \\
5:1     &   2.924    &88.01  & 0.3146  &  3.5554 & 0.1473 & 2.5490 \\
11:2    &   3.116    &93.79  & 0.2411  &  3.7966 & 0.1107 & 2.6597 \\
6:1     &   3.302    &99.39  & 0.3027  &  4.0993 & 0.1303 & 2.7900 \\
13:2    &   3.483    &104.84 & 0.2360  &  4.3353 & 0.1033 & 2.8933 \\
7:1     &   3.659    &110.15 & 0.2947  &  4.6301 & 0.1158 & 3.0090 \\
15:2    &   3.832    &115.33 & 0.2244  &  4.8545 & 0.0852 & 3.0942 \\
8:1     &   4.000    &120.40 & 0.2806  &  5.1351 & 0.1093 & 3.2036 \\
17:2    &   4.165    &125.37 & 0.2116  &  5.3467 & 0.0764 & 3.2800 \\
9:1     &   4.327    &130.24 & 0.2611  &  5.6077 & 0.0933 & 3.3733 \\
19:2    &   4.486    &135.02 & 0.2037  &  5.8114 & 0.0673 & 3.4406 \\
10:1    &   4.642    &139.71 & 0.2481  &  6.0595 & 0.0851 & 3.5257 \\
\hline
\end{tabular}
\end{center}
\label{t:area}
\end{table}

\begin{figure}
\gridline{\fig{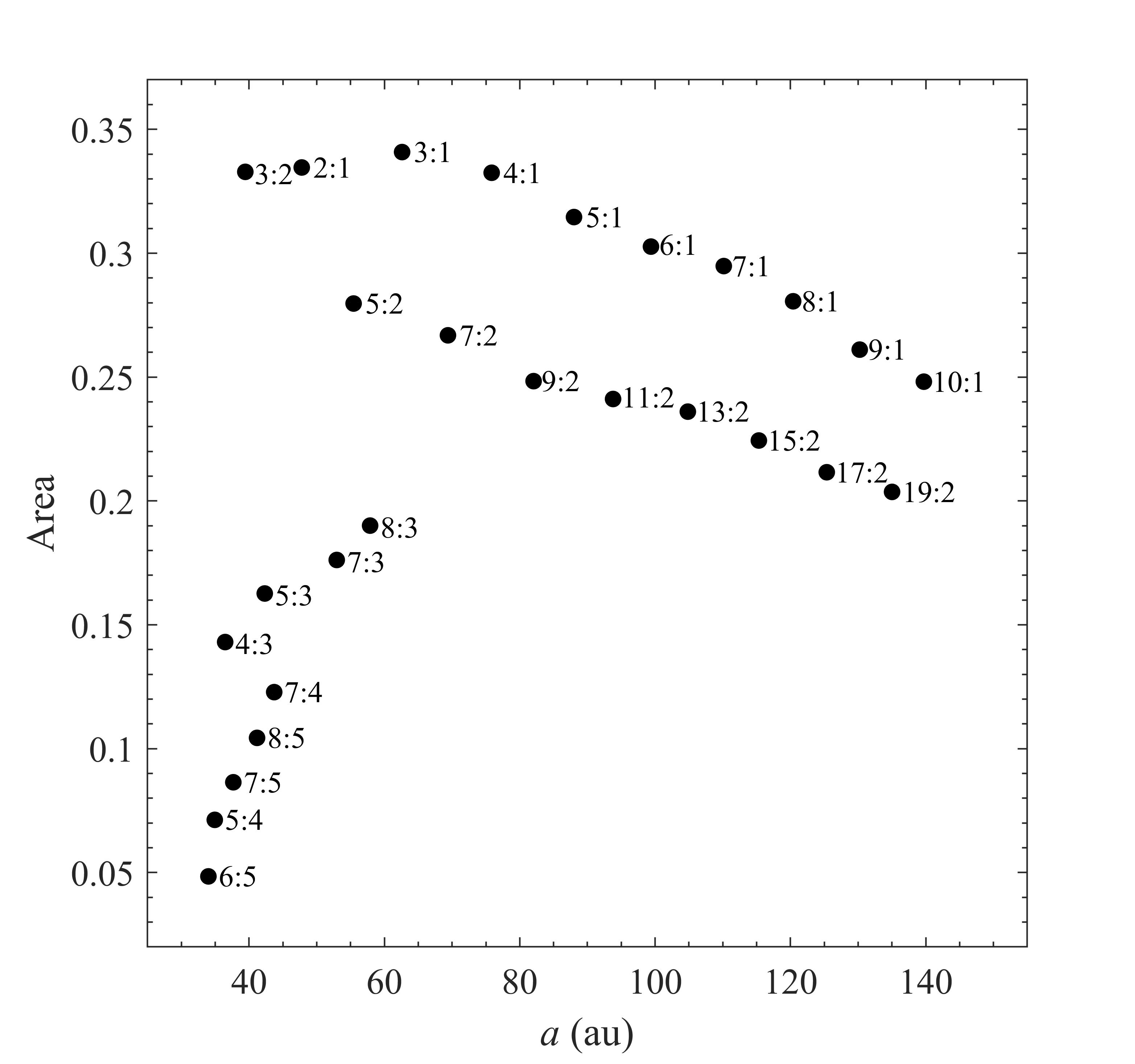}{0.5\textwidth}{(a)}
          \fig{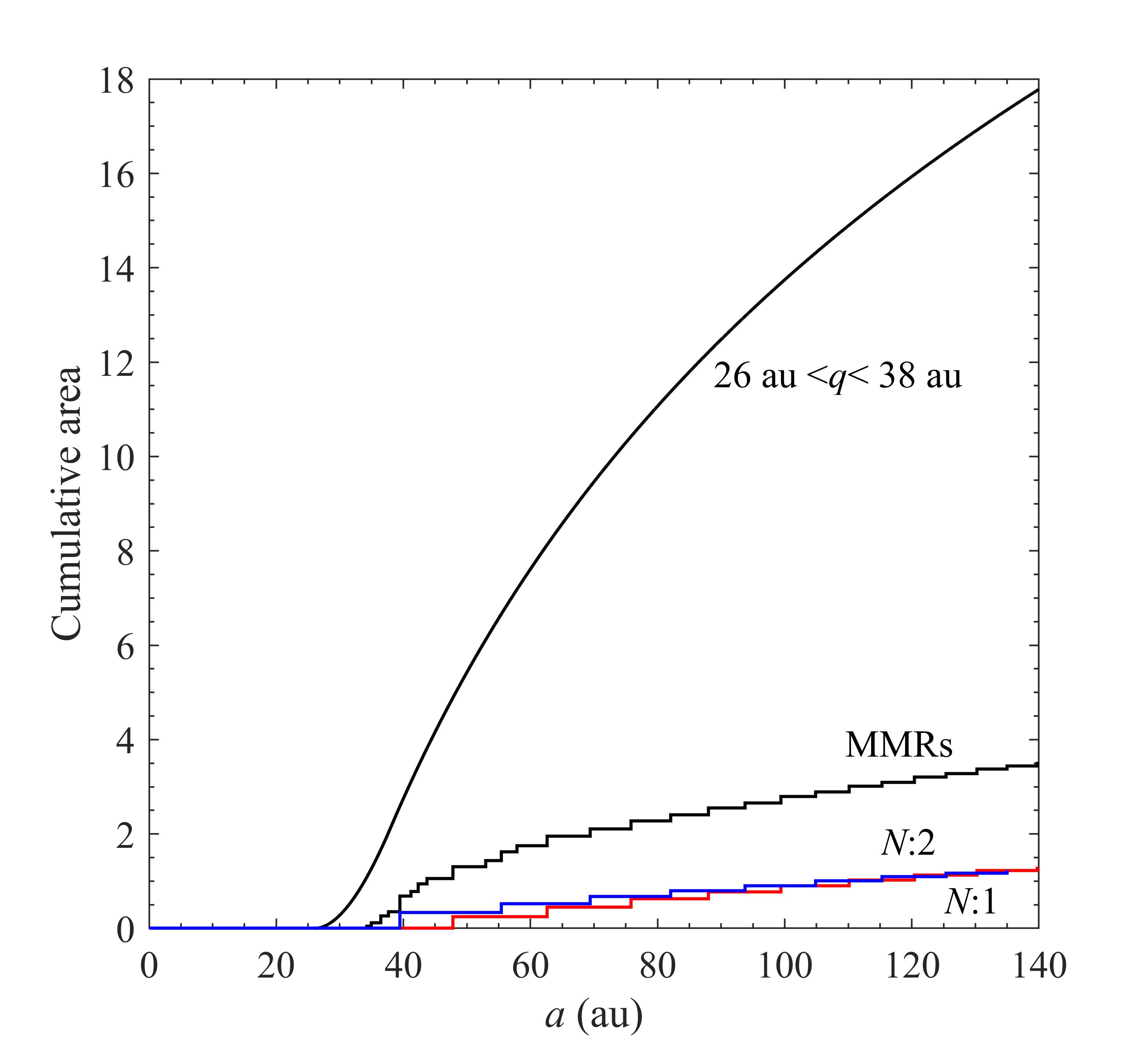}{0.5\textwidth}{(b)}
          }
\caption{
(a) The area, in the $(a,e)$ plane bounded by perihelion distance $q>26$ au, of the first resonance zones of Neptune's exterior MMRs.
(b) The cumulative area, in the $(a,e)$ plane bounded by the perihelion distance range 26~au~$<q<38$~au, of the first resonance zones of MMRs (lower curves), and the total cumulative area (top curve). (The areas are in units of au.)
}
 \label{f:areas}
\end{figure}

\subsubsection{A large fraction of the SDOs' phase space is stable resonant zones}\label{subsubsec:stablefraction}
Because ``resonance sticking" in Neptune's exterior MMRs appears to be the main mechanism for lengthening the dynamical lifetimes of scattered disk objects~\citep{Duncan:1997,Lykawka:2007b,Yu:2018}, it is useful to measure the sizes of the resonance zone boundaries. 
As a rough indication of the sizes of resonance zone boundaries, we measured the area of the stable resonance zones in the $(a,e)$ plane for all the MMRs that we investigated. In measuring the resonant areas, we adopted a cutoff perihelion distance $q>26$ au because Neptune-resonant orbits of lower perihelion distance are subject to destabilizing perturbations from the other giant planets (Uranus, Saturn, etc.) on timescales of less than 10 megayears~\citep[e.g.][]{Malhotra:2018}. Within this perihelion cutoff, most MMRs have only a small segment of the second resonance zone therefore we focussed on measuring the area of the first resonance zone. The measured resonant areas are listed in Table~\ref{t:area} and also plotted in Figure~\ref{f:areas}. We also measured and tabulated the resonant areas in the sub-region bounded by 26~au~$<q<38$~au; the upper boundary of this perihelion distance range is chosen to generously cover the fuzzy boundary of SDOs noted above.  

The measured cumulative resonant area (Figure~\ref{f:areas}b) is found to be approximately 22\% of the entire area in the $(a,e)$ plane bounded by 26~au~$<q<38$~au and 33~au~$<a<140$~au. This fraction should be considered a lower limit because we have not measured many resonances with particle-to-Neptune period ratios in-between the resonances that we have measured in our study nor have we measured the more distant resonances at $a>140$ au. 
This implies that in the orbital parameter regime occupied by the SDOs, the long term stable resonant zones occupy a large fraction of the phase space, even at large semi-major axes, explaining the prominence of the resonance sticking behavior of SDOs in numerical simulations.  

Examining the tabulated results, we observe that, individually, the 2:1 and 3:2 resonances have very similar area in the $(a,e)$ plane, the $N$:1 sequence of MMRs have the largest areas, and the resonant areas decrease rather slowly with increasing $N$. The areas of the $N$:2 resonances are comparable to but lower than those of the nearby $N$:1 resonances; these also decrease rather slowly with increasing period ratio. These trends confirm
 the general result from numerical simulations that the $N$:1 resonances are the strongest/stickiest in the scattered disk, followed by the $N$:2 resonances \citep{Lykawka:2007b,Yu:2018}.  

 We can compare in more detail the trends of the $(a,e)$ resonant areas with the trends in ``resonance strength" computed in \citet[Figure 7]{Gallardo:2006b} who defined ``resonance strength" as $SR(a,e,i,\omega) = \langle R\rangle - R_{\rm min}$, where $i$ and $\omega$ are the orbital inclination and the argument of perihelion, $\langle R\rangle$ is the orbit-averaged value of the resonant disturbing function $R$, and $R_{\rm min}$ is the minimum value of $R$. The author reported the strengths of many of Neptune's exterior MMRs for semi-major axes up to $\sim300$~au, for fixed values of $\omega=60^\circ$, $i=20^\circ$ and perihelion distance 32~au. The general trend of decreasing resonant area with increasing semi-major axis is consistent with the trends reported in ``resonance strength", but there are some quantitative differences. In our results, the area of the 3:1 resonance is slightly larger than that of the 2:1 resonance and the area of the 2:1 resonance is slightly larger than that of the 3:2 resonance; \cite{Gallardo:2006b} finds these reversed in resonance strength. At larger semi-major axis, we find that the resonant areas decrease more slowly than the faster decrease of resonance strength. For example, within the SDOs' approximate perihelion distance range of $26~{\rm au}<q<38~{\rm au}$, the area of the 9:1 resonance is about 38\% that of the 2:1 resonance, whereas the resonance strength of the 9:1 is about 10\% that of the 2:1.  

We can also compare our results on the $(a,e)$ resonant areas with the relative ``resonance stickiness" computed in \citet[their Figure 5]{Lykawka:2007a} who defined ``resonance stickiness" as the fraction of time spent in each resonance by the scattered disk test particles that survived to 4 Gyr in their numerical simulations. The general trend is that resonance stickiness of $N$:1 and $N$:2 resonances decreases with increasing semi-major axis which agrees with our results, although the 6:1 resonance has the largest reported stickiness and the 11:1 also has an anomalously large stickiness compared to adjacent resonances; these anomalies are likely due to small number statistics of the surviving test particles in the simulations. Overall, the resonant areas in the $N$:1 and $N$:2 sequences decrease more slowly with $N$ than resonance stickiness (and the resonance stickiness decreases more slowly than resonance strength). For example, the resonant area of the 9:1 MMR is approximately 38\% that of the 2:1, compared to the resonance stickiness ratio of 20\% (and compared to 10\% for the resonance strengths).  \cite{Lykawka:2007a} also reported that the scattered disk test particles that survived to 4 Gyr spent $\sim38\%$ of their lifetimes sticking in resonances and that $\sim33\%$ of the test particles were found in resonances at the end of the simulation; these fractions are consistent with the 22\% lower limit for the fractional area represented by resonances that we found.

\subsubsection{A tool to identify candidate resonant objects}\label{subsubsec:tool}
The current method of identifying and classifying observed resonant objects (including resonance sticking SDOs) involves arduous visual examination of the libration behavior of the critical resonant angles in 10 mega year long numerical integrations with a solar system model that includes the perturbations of the giant planets~\citep[e.g.][]{Gladman:2008,Volk:2016}. 
The boundaries of Neptune's resonances in the $(a,e)$ plane that we have computed here can be used to identify candidate SDOs from knowledge of just their observed orbital elements, specifically, their barycentric semi-major axis and eccentricity, thus greatly reducing the computational and human effort. The data for the resonance boundaries in the $(a,e)$ plane are available in electronic form upon request. In the future, this tool can be made more useful by computing the resonance boundaries of a more complete set of Neptune's resonances than we have attempted here.

\bigskip
{
$\mathbf{Acknowledgements}$: We thank Kathryn Volk for discussions.
 We also thank Tabare Gallardo and an anonymous reviewer for providing helpful reviews.
LL acknowledges funding from National Natural Science Foundation of China (11572166) and China Scholarship Council.
RM acknowledges funding from NASA (grants NNX14AG93G and 80NSSC19K0785) and NSF (grant AST-1824869).
}

\end{document}